\documentclass[twocolumn,showpacs,amsmath,amssymb]{revtex4}
\usepackage{graphicx}% Include figure files
\usepackage{dcolumn}% Align table columns on decimal point
\usepackage{bm}% bold math

\newcommand{\glo}{g_{\|}}
\newcommand{\gtr}{g_{\perp}}
\newcommand{\D}{{\rm d}}
\newcommand{\Delm}{\Delta_{m}}
\newcommand{\Hlo}{H_{\|}}
\newcommand{\Htr}{H_{\perp}}
\newcommand{\e}{\mathrm{e}}

\newcommand{\iu}{{\rm i}}
\newcommand{\K}{D}
\newcommand{\kT}{k_{{\rm B}}T}

\newcommand{\w}{\omega}
\newcommand{\tdec}{\tau_{\rm dec}}
\newcommand{\tphi}{\tau_{\Phi}}

\newcommand{\rate}{\Gamma}%{\Gamma(H=0)}
\newcommand{\rateo}{\Gamma_{0}}%{\Gamma(H=0)}

\newcommand{\cth}{\mathrm{cth}\,}

\newcommand{\half}{\tfrac{1}{2}}

\newcommand{\Mz}{M_{z}}

\newcommand{\llangle}{\left\langle}
\newcommand{\rrangle}{\right\rangle}

\newcommand{\el}{\varepsilon} % Energy levels

\newcommand{\Z}{{\cal Z}}

\newcommand{\Xnl}{\chi_{3}}
\newcommand{\Xnlw}{\chi_{3}^{(\w)}}
\newcommand{\Xnlo}{\chi_{3}^{(3\w)}}

\newcommand{\Xnlwt}{\tilde{\chi}_{3}}

\newcommand{\dBlo}{b_{\|}}
\newcommand{\dBtr}{b_{\perp}}

\newcommand{\Dchi}{\Delta\chi}

\newcommand{\QF}{\bar{Q}}
\newcommand{\QFt}{Q}

\begin{document}

%\bibliographystyle{prsty}%{apsrev}

%%%%%%%%%%%%%%%%%%%%%%%%%%%%%%%%%%%%%%%%%%%%%%%%%%%%%%%%%%%%
%%%%%%%%%%%%%%%%%%%%%%%%%%%%%%%%%%%%%%%%%%%%%%%%%%%%%%%%%%%%
%%%%%%%%%%%%%%%%%%%%%%%%%%%%%%%%%%%%%%%%%%%%%%%%%%%%%%%%%%%%
%%%%%%%%%%%%%%%%%%%%%%%%%%%%%%%%%%%%%%%%%%%%%%%%%%%%%%%%%%%%
%%%%%%%%%%%%%%%%%%%%%%%%%%%%%%%%%%%%%%%%%%%%%%%%%%%%%%%%%%%%

\title{
Nonlinear response of single-molecule nanomagnets:
equilibrium and dynamical
}

\author{
R.~L\'{o}pez-Ruiz, F. Luis,
V. Gonz\'{a}lez, A. Mill\'an,
and J.~L. Garc\'{\i}a-Palacios
}
\affiliation{
Instituto de Ciencia de Materiales de Arag\'on y
Dep.\ de F\'{\i}sica de la Materia Condensada,
C.S.I.C. -- Universidad de Zaragoza, E-50009 Zaragoza, Spain
}

%%%%%%%%%%%%%%%%%%%%%%%%%%%%%%%%%%%%%%%%%%%%%%%%%%%%%%%%

\date{\today}

\begin{abstract}
We present an experimental study of the {\em nonlinear\/}
susceptibility of Mn$_{12}$ single-molecule magnets.
We investigate both their thermal-equilibrium and dynamical nonlinear
responses.
The equilibrium results show the sensitivity of the nonlinear
susceptibility to the magnetic anisotropy, which is nearly absent in
the linear response for axes distributed at random.
The nonlinear dynamic response of Mn$_{12}$ was recently found to be
very large and displaying peaks reversed with respect to classical
superparamagnets [F.~Luis {\em et al.}, Phys.\ Rev.\ Lett.\ {\bf 92},
107201 (2004)].
Here we corroborate the proposed explanation --- strong field
dependence of the relaxation rate due to the detuning of tunnel energy
levels.
This is done by studying the orientational dependence of the nonlinear
susceptibility, which permits to isolate the quantum detuning
contribution.
Besides, from the analysis of the longitudinal and transverse
contributions we estimate a bound for the decoherence time due to the
coupling to the phonon bath.
\end{abstract}

\pacs{75.50.Xx, 75.50.Tt, 75.45.+j, 75.40.Gb}

\maketitle

%\tableofcontents

%%%%%%%%%%%%%%%%%%%%%%%%%%%%%%%%%%%%%%%%%%%%%%%%%%%%%%%%%%%%
%%%%%%%%%%%%%%%%%%%%   INTRODUCTION   %%%%%%%%%%%%%%%%%%%%%%
%%%%%%%%%%%%%%%%%%%%   INTRODUCTION   %%%%%%%%%%%%%%%%%%%%%%
%%%%%%%%%%%%%%%%%%%%   INTRODUCTION   %%%%%%%%%%%%%%%%%%%%%%
%%%%%%%%%%%%%%%%%%%%   INTRODUCTION   %%%%%%%%%%%%%%%%%%%%%%
%%%%%%%%%%%%%%%%%%%%   INTRODUCTION   %%%%%%%%%%%%%%%%%%%%%%
%%%%%%%%%%%%%%%%%%%%%%%%%%%%%%%%%%%%%%%%%%%%%%%%%%%%%%%%%%%%

\section{
Introduction
}

Superparamagnets are nanoscale solids or clusters with a large net
spin ($S\sim10^{1}$--$10^{4}$).
This spin is coupled to the environmental degrees of freedom of the
host material, e.g., phonons, nuclear spins, or conducting electrons.
Due to the dynamical disturbances of the surroundings the spin may,
among other things, undergo a Brownian-type ``reversal'', overcoming
the potential barriers created by the magnetic anisotropy.

Depending on the relation between the reversal time $\tau$ and the
observation time $t_{\rm obs}$, different phenomenologies can be
found.
For $\tau\ll t_{\rm obs}$, the spin exhibits the thermal-equilibrium
distribution of orientations as in a paramagnet; the large values of
$S$ are the reason for the name {\em superparamagnetism}.
When $\tau\gg t_{\rm obs}$, in contrast, the reversal mechanisms
appear {\em blocked\/} and the spin stays close to an energy minimum
(stable magnetization conditions appropriate for magnetic storage).
Finally, under intermediate conditions ($\tau\sim t_{\rm obs}$) one
finds {\em non-equilibrium phenomena\/} (i.e., magnetic
``relaxation").

\begin{figure}[b]
%\hspace*{-2.em}
\resizebox{7.5cm}{!}{\includegraphics{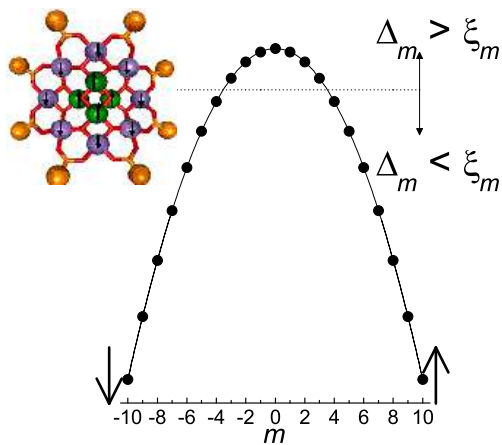}}
\vspace*{-5.ex}
\caption{
Energy levels for Mn$_{12}$ (the molecule is sketched in the inset
along with the spin orientations of the Mn ions).
The levels are plotted vs.\ the quantum number $m$ at $H=0$ and show
the bistable potential for the spin due to the magnetic anisotropy.
The horizontal line marks the border between ``classical'' or
localized energy levels, $\Delta_{m}<\xi_{m}$, and the tunneling
levels $\Delta_{m}>\xi_{m}$ ($\Delta$ is the tunnel splitting and
$\xi$ the width of the environmental bias-field distribution).
}
\vspace*{-3.ex}
\label{fig:levels}
\end{figure}

For large $S$, for instance in magnetic nanoparticles \cite{panpol93},
a classical description is adequate \cite{bro63}.
The essential physical ingredients are the thermo-activation over the
magnetic anisotropy barriers and the (damped) spin precession
\cite{endnote41}.
As the spin value is reduced, quantum effects can start to play a role.
For moderate spins ($S\sim10$), as in single-molecule magnets
\cite{harpolvil96}, the quantum nature of the system comes
significantly to the fore.
For instance, the spin reversal may also occur by tunneling whenever
the magnetic field brings into resonance quantum states located at the
sides of the barrier (Fig.~\ref{fig:levels}).
%_____________________________
%_____________________________
%_____________________________
%_____________________________
%_____________________________
%_____________________________

Several fundamental problems can be studied on these systems
\cite{zur91,leg2002,weiss}.
First, the quantum-to-classical transition, with the emergence of
classical properties.
Single-molecule magnets constitute a model system to study quantum
mechanics at a mesoscopic level, while magnetic nanoparticles provide
a natural classical limit.
Second, one can address the effects of environmental degrees of
freedom on a given system.
Classically, one faces the rich phenomenology of rotational Brownian
motion of the nanoparticle magnetization.
In the quantum case the bath coupling not only produces fluctuations
and dissipation (allowing the system to relax to thermal equilibrium),
but it is also responsible for the decoherence of its quantum
dynamics.
Thus, single-molecule magnets constitute an important experimental
benchmark to test the predictions of the theory of {\em quantum
dissipative systems\/} \cite{weiss} (much as Josephson junctions in
superconductivity).

The best studied magnetic molecular clusters are those named Mn$_{12}$
and Fe$_{8}$, both with $S=10$ in the ground multiplet (for other
examples see Ref.~\cite{blupra2004}).
To understand their behavior a plethora of theoretical calculations
and every conceivable experimental technique have been used.
However, the nonlinear susceptibility $\Xnl$, fruitfully exploited in
studies of spin glasses \cite{wuetal93,hegkiskem94,schetal95}, random
anisotropy systems \cite{RAM}, but also in magnetic nanoparticles
\cite{bitetal93,jonsvehan98,jonjongarsve2000}, had been overlooked.
For {\em classical\/} superparamagnets $\Xnl$ provides information on
parameters to which the linear susceptibility is less sensitive, like
the anisotropy constant $\K$ \cite{garjonsve2000} or the spin-bath
coupling parameter $\lambda$ \cite{garsve2000} (which enters the scene
due to the strong dependence of the relaxation rate $\rate=1/\tau$ on
{\em transverse\/} magnetic fields \cite{garkencrocof99}).
Besides, the dynamical nonlinear susceptibility has a genuinely {\em
quantum\/} contribution due to the detuning of the energy levels by a
{\em longitudinal\/} field \cite{luietal2004a}.
It has a sign opposite to the classical (precessional) contribution,
thus allowing to ascertain whether quantum effects, such as resonant
tunneling, are relevant in a given nanomagnet (an issue sometimes
controversial \cite{mamnakfur2002}).

In this article we present experimental results for the
thermal-equilibrium and dynamical nonlinear susceptibility of
Mn$_{12}$ acetate.
Compared to the linear susceptibility, the equilibrium $\Xnl$ shows an
enhanced sensitivity to the magnetic anisotropy, even for axes
distributed at random (allowing to estimate $\K$ from measurements in
powdered samples).
%
%It also reflects quantum effects due to the discrete nature of the
%energy levels.
%
As for the frequency-dependent $\Xnl$, we study its dependence on the
angle of the applied field.
This gives direct access to the relaxation-rate field-expansion
coefficients,
$\rate-\rate|_{H=0}
\propto
\glo\Hlo^{2}+\gtr\Htr^{2}$,
which contain valuable information on the spin reversal mechanisms,
thus allowing to separate the ``classical-transverse'' and
``quantum-longitudinal'' contributions to $\Xnl$.

In the discussion simple approximate formulas and numerical results
from the solution of a Pauli quantum-master equation are used.
Our investigation confirms the interpretation of
Ref.~\cite{luietal2004a} of the large quantum contribution to
$\Xnl(\w)$ as arising from the detuning of the tunnel channels by a
longitudinal magnetic field.
Thus, the experimental nonlinear response is consistent with the
established scenario
\cite{frietal96,heratal96,thoetal96,garchu97,luibarfer98,wur98} of
thermally-activated tunnel via excited states in Mn$_{12}$.
The analysis also gives a bound for the decoherence time $\tphi$
(timescale for the attainment of a diagonal density matrix due to the
coupling to the phonon bath).
The obtained $\tphi$ turns out to be much shorter than the lifetime of
the spin levels $\tau_{0}$, and is responsible for a fast loss of
coherent dynamics (like tunnel oscillations or precession).

%%%%%%%%%%%%%%%%%%%%%%%%%%%%%%%%%%%%%%%%%%%%%%%%%%%%%%%%%%%%
%%%%%%%%%%%%%%%%%   EXPERIMENTAL  DETAILS   %%%%%%%%%%%%%%%%
%%%%%%%%%%%%%%%%%   EXPERIMENTAL  DETAILS   %%%%%%%%%%%%%%%%
%%%%%%%%%%%%%%%%%   EXPERIMENTAL  DETAILS   %%%%%%%%%%%%%%%%
%%%%%%%%%%%%%%%%%   EXPERIMENTAL  DETAILS   %%%%%%%%%%%%%%%%
%%%%%%%%%%%%%%%%%   EXPERIMENTAL  DETAILS   %%%%%%%%%%%%%%%%
%%%%%%%%%%%%%%%%%%%%%%%%%%%%%%%%%%%%%%%%%%%%%%%%%%%%%%%%%%%%

\section{
Samples and measurements
}
\label{sec:samples-measurements}

%%%%%%%%%%%%%%%%%%%%%%%%%%%%%%%%%%%%%%%%%%%%%%%%%%%%%%%%%%%%
%%%%%%%%%%%%%%%%%%%%%%%%   SAMPLES   %%%%%%%%%%%%%%%%%%%%%%%
%%%%%%%%%%%%%%%%%%%%%%%%   SAMPLES   %%%%%%%%%%%%%%%%%%%%%%%
%%%%%%%%%%%%%%%%%%%%%%%%   SAMPLES   %%%%%%%%%%%%%%%%%%%%%%%
%%%%%%%%%%%%%%%%%%%%%%%%%%%%%%%%%%%%%%%%%%%%%%%%%%%%%%%%%%%%

\subsection{
Samples and set up
}

Single crystals of Mn$_{12}$ acetate were grown following a procedure
similar to that described by Lis \cite{lis80}.
The concentrations of the reactants, however, were higher than those
of \cite{lis80} in order to increase the supersaturation and the
growth rate, yielding larger crystals.
These were regrown several times by renewing the mother solution.
X-ray diffraction patterns of powdered crystals agreed with simulated
patterns from the known crystal structure.

The measurements were performed on a single crystal with dimensions
$3\times0.5\times0.5$\,mm$^{3}$ at different orientations with respect
to the applied magnetic field.
To this end, we constructed a rotating sample holder that enables the
$c$ crystallographic axis (which defines the anisotropy axes of the
Mn$_{12}$ molecules) to be rotated a given angle $\psi$ with respect
to the magnet axis.
This angle is measured with a precision better than $0.5$ degrees.
To calibrate the position of the zero we used the measured linear
equilibrium susceptibility, which should be maximum when the field is
parallel to the anisotropy axis ($\psi=0$).
%_____________________________
%_____________________________
%_____________________________
\begin{figure}
\hspace*{-2.em}
\resizebox{9.cm}{!}{\includegraphics{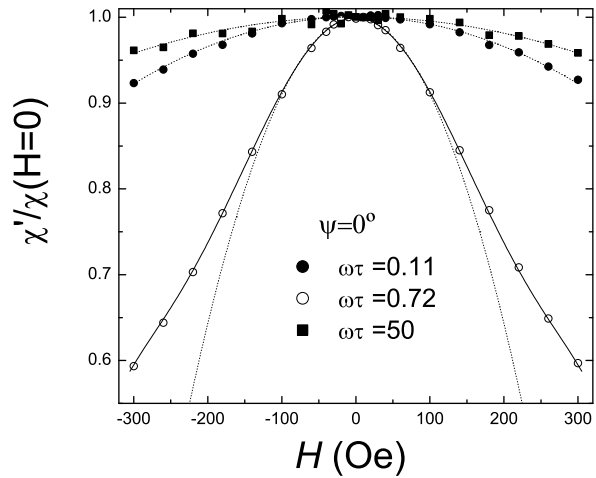}}
\vspace*{-5.ex}
\caption{
Magnetic ac susceptibility of a single crystal of Mn$_{12}$,
normalised by its zero-field value, vs.\ the static field $H$
(parallel to the anisotropy axes, $\psi=0$).
Results for the {\em real part\/} at $T=5$\,K and various frequencies
$\w\tau$ are shown.
The solid lines represent polynomial fits from which $\Xnl$ is
obtained.
The parabolic approximation $\chi_{1} + 3\Xnl H^{2}$, which dominates
the low-field behavior ($|H|\lesssim100$\,Oe), is shown by dotted
lines.
}
\vspace*{-3.ex}
\label{XvsH}
\end{figure}
%_____________________________
%_____________________________
%_____________________________

%%%%%%%%%%%%%%%%%%%%%%%%%%%%%%%%%%%%%%%%%%%%%%%%%%%%%%%%%%%%
%%%%%%%%%%%%%%%%   MAGNETIC  MEASUREMENTS   %%%%%%%%%%%%%%%%
%%%%%%%%%%%%%%%%   MAGNETIC  MEASUREMENTS   %%%%%%%%%%%%%%%%
%%%%%%%%%%%%%%%%   MAGNETIC  MEASUREMENTS   %%%%%%%%%%%%%%%%
%%%%%%%%%%%%%%%%%%%%%%%%%%%%%%%%%%%%%%%%%%%%%%%%%%%%%%%%%%%%

\subsection{
Magnetic measurements
}

Dynamical susceptibility measurements were performed using the ac
option of a commercial SQUID magnetometer, by applying an alternating
field $\sim\Delta h\,{\rm e}^{\iu\,\w t}$.
The ac susceptibility was measured under a weak superimposed dc field
$H$, parallel to the oscillating one.
The first harmonic of the response can then be expanded as
%_____________________________
%
% EXPANSION DYNAMICAL SUSCEPTIBILITY
%_____________________________
\begin{equation}
\chi(\w,H)
=
\chi_{1}(\w)
+
3\Xnl(\w)H^{2}
+
5\chi_{5}(\w)H^{4}
+
\cdots
\;.
\label{expansion}
\end{equation}
%_____________________________
%_____________________________
The ($H$-independent) expansion coefficients give the ordinary linear
susceptibility $\chi_{1}$ and the nonlinear ones $\Xnl$,
$\chi_{5}$,\dots.
As it is customary, we focus on $\Xnl$ and refer to it as {\em the\/}
nonlinear susceptibility.

To determine the nonlinear susceptibility we performed polynomial fits
of the $H$-dependent ac data whose quadratic coefficient gives $\Xnl$.
An illustrative example of the fitting procedure is shown in
Fig.~\ref{XvsH}.
For sufficiently low $H$, a good description is provided by a simple
parabolic dependence $\chi_{1} + 3\Xnl H^{2}$.
For increasingly larger fields, we increased the order of the
polynomials whenever the fitting error became greater than $5 \%$.
The experimental $\Xnl$ was taken as the mean value of all quadratic
coefficients obtained from the different order polynomials, thus
miminizing the error of the determination.
%_____________________________
%_____________________________
%_____________________________
\begin{figure}
\hspace*{-2.em}
\resizebox{9.cm}{!}{\includegraphics{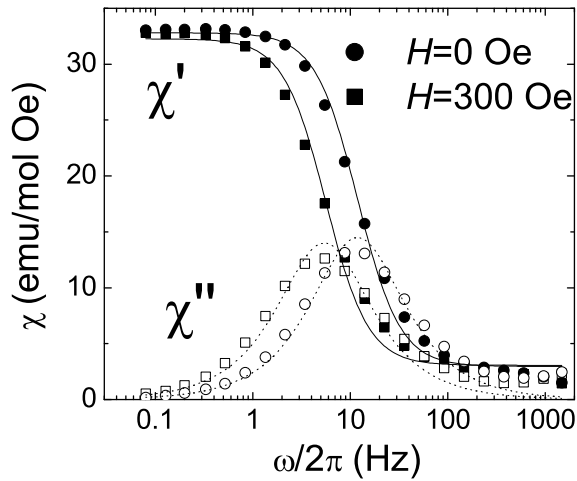}}
\vspace*{-5.ex}
\caption{
Magnetic susceptibility vs.\ frequency measured along the anisotropy
axis ($\psi=0$) at $T=5$\,K.
Results are shown at zero bias field (circles) and at $H=300$\,Oe
(squares).
Solid symbols are for the real part (in-phase component) and open
symbols for the imaginary part (out-of-phase).
The lines are fits to the Debye law (\ref{debye}) with
$\tau|_{H=0}=1.3(3)\times10^{-2}$\,s and
$\tau|_{H=300}=2.8(1)\times10^{-2}$\,s.
%
%(0.013272401760267=3.e-8*exp(65./5.))
%
}
\vspace*{-3.ex}
\label{X1vsfreq}
\end{figure}
%_____________________________
%_____________________________
%_____________________________

The measurements were performed at temperatures in the range
$2$\,K${}<T<45$\,K and frequencies $0.08$\,Hz${}<\w/2\pi<1.5$\,kHz.
The amplitude of the ac field ($\Delta h=3$\,Oe) was sufficiently
small not to induce nonlinearity in $\Delta h$ (associated to the
generation of harmonics).
Actually, the $\Xnl(\w)$ obtained here from the field-dependent
susceptibility differs from the $\Xnl$ extracted from the third
harmonic of the response to an ac field (at $H=0$).
However, as we show in Appendix~\ref{app:analytic}, the dependence of
both quantities on the parameters under study is analogue.

%%%%%%%%%%%%%%%%%%%%%%%%%%%%%%%%%%%%%%%%%%%%%%%%%%%%%%%%%%%%
%%%%%%%%%%%%%%%%   EQ & DYN  MEASUREMENTS   %%%%%%%%%%%%%%%%
%%%%%%%%%%%%%%%%   EQ & DYN  MEASUREMENTS   %%%%%%%%%%%%%%%%
%%%%%%%%%%%%%%%%   EQ & DYN  MEASUREMENTS   %%%%%%%%%%%%%%%%
%%%%%%%%%%%%%%%%%%%%%%%%%%%%%%%%%%%%%%%%%%%%%%%%%%%%%%%%%%%%

\subsection{
Equilibrium vs.\ dynamical measurements
}

In the next sections we are going to study the equilibrium and
dynamical susceptibilities.
Let us define here a practical criterion to decide when the
experimentally measured $\chi_{1}(\w)$ and $\Xnl(\w)$ correspond to
equilibrium or off-equilibrium conditions.
In the temperature range covered by our experiments, $T>2$\,K, the
molecular spins of Mn$_{12}$ relax via a thermally-activated tunneling
mechanism \cite{frietal96,heratal96,thoetal96}.
This process gives rise to a well-defined relaxation time $\tau$ and
the ac response can be described by a simple Debye formula:
%_____________________________
%
% DEBYE LAW
%_____________________________
\begin{equation}
\chi
=
\chi_{S}
+
\frac{\chi_{T}-\chi_{S}}{1+\iu\,\w\tau}
\;.
\label{debye}
\end{equation}
%_____________________________
%_____________________________
Here $\chi_{T}$ and $\chi_{S}$ are the isothermal (thermal
equilibrium) and adiabatic limits of $\chi$.
In Fig.~\ref{X1vsfreq} we show how the response of the Mn$_{12}$
crystals follows Eq.~(\ref{debye}) at temperatures and magnetic fields
typical of our experiments \cite{endnote42}.
The equilibrium regime corresponds to frequencies fulfilling
$\w\tau\ll 1$ (left part of the plot), relaxation effects becoming
important when the range $\w\tau\sim1$ is approached.
%_____________________________
%_____________________________
%_____________________________
\begin{figure}
\hspace*{-2.em}
\resizebox{9.cm}{!}{\includegraphics{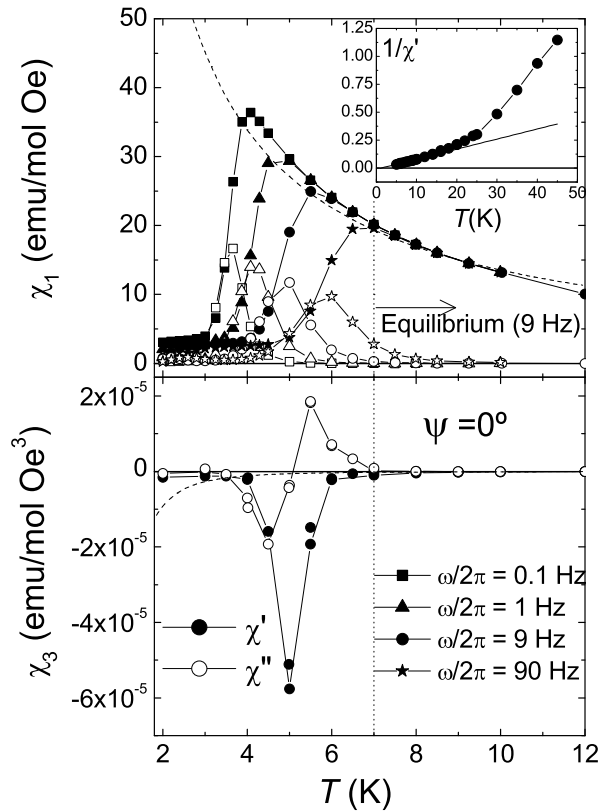}}
\vspace*{-5.ex}
\caption{
Linear (top panel) and nonlinear susceptibilities (bottom) vs.\
temperature for several frequencies (only $9$\,Hz for $\Xnl$) measured
along the anisotropy axis.
Solid symbols are for the real parts and open symbols for the
imaginary parts.
Dashed lines are the equilibrium susceptibilities in the Ising (large
anisotropy) limit [Eqs.~(\ref{Ising1}) and (\ref{Ising2})].
The vertical line marks the boundary above which thermal equilibrium
results are safely obtained using $\w/2\pi=9$\,Hz.
Inset: temperature dependence of the reciprocal {\em equilibrium\/}
linear susceptibility and its fit to a Curie-Weiss law
(\ref{Curie-Weiss}) for $T<10$\,K.
}
\vspace*{-3.ex}
\label{XvsT}
\end{figure}
%_____________________________
%_____________________________
%_____________________________

On the other hand, the relaxation time of this system increases
exponentially as $T$ decreases, following an Arrhenius law (see for
instance Fig.~2 in Ref.~\cite{luietal2004a})
%_____________________________
%
% ARRHENIUS LAW
%_____________________________
\begin{equation}
\tau
=
\tau_{0}
\exp{(U/\kT)}
\;.
\label{arrhenius}
\end{equation}
%_____________________________
%_____________________________
Here $U$ is an activation energy and $\tau_{0}$ an attempt time, which
set the magnitude and temperature dependence of $\tau$.
At zero field we obtained $U_{0}\simeq 65$\,K and
$\tau_{0}\simeq3\times10^{-8}$\,s \cite{luietal2004a}.
In a $\chi$ vs.\ $T$ experiment (Fig.~\ref{XvsT}) the condition
$\w\tau=1$ defines a superparamagnetic ``blocking" temperature
$\kT_{\rm b}
=
-U_{0}\ln(\w\tau_{0})$.
Below $T_{\rm b}$, the real part $\chi_{1}'$ drops from $\chi_{T}$
towards $\chi_{S}$ whereas $\chi_{1}''$ departs from zero and shows a
maximum.
As for the nonlinear susceptibility, the non-equilibrium response near
$T_{\rm b}$ leads also to a non-zero imaginary part $\Xnl''$, and to a
strong deviation of $\Xnl'$ from the equilibrium $\chi_{3 T}$.
In contrast with the linear response (which decreases from
$\chi_{T}$), $\Xnl'$ becomes {\em larger\/} than $\chi_{3 T}$ near
$T_{\rm b}$ \cite{luietal2004a}; this dynamical nonlinear phenomenon
will be discussed in detail in the following sections.

As can be seen in Figs.~\ref{X1vsfreq} and \ref{XvsT}, the transition
from isothermal to adiabatic conditions extends over a certain
frequency or temperature range, determined by the width of the $\chi''
(\w,T)$ curve.
As a practical rule, we take $\chi_{T}=\chi_{1}'$ and $\chi_{3T}=\Xnl'$
when the imaginary parts are reasonably small $\chi''/\chi'<10^{-2}$.
This value is, as a matter of fact, close to the experimental error of
the present measurements.
Using this criterion we have extracted, from the dynamical
susceptibility data vs.\ $T$, the equilibrium $\chi_{T}$ and
$\chi_{3T}$ discussed in the next section.

%%%%%%%%%%%%%%%%%%%%%%%%%%%%%%%%%%%%%%%%%%%%%%%%%%%%%%%%%%%%
%%%%%%%%%%%%%%%%%   EQUILIBRIUM  RESPONSE   %%%%%%%%%%%%%%%%
%%%%%%%%%%%%%%%%%   EQUILIBRIUM  RESPONSE   %%%%%%%%%%%%%%%%
%%%%%%%%%%%%%%%%%   EQUILIBRIUM  RESPONSE   %%%%%%%%%%%%%%%%
%%%%%%%%%%%%%%%%%   EQUILIBRIUM  RESPONSE   %%%%%%%%%%%%%%%%
%%%%%%%%%%%%%%%%%   EQUILIBRIUM  RESPONSE   %%%%%%%%%%%%%%%%
%%%%%%%%%%%%%%%%%%%%%%%%%%%%%%%%%%%%%%%%%%%%%%%%%%%%%%%%%%%%

\section{
Equilibrium linear and nonlinear susceptibilities
}
\label{sec:equilibrium}

The equilibrium $\chi_{T}$ obtained as described in the previous
section is shown in the inset of Fig.~\ref{XvsT}.
In the temperature range $4$\,K${}<T<10$\,K it approximately follows a
Curie-Weiss law
%_____________________________
%
% CURIE-WEISS LAW
%_____________________________
\begin{equation}
\chi_{T}
=
\frac{C}{T-\theta}
\;,
\label{Curie-Weiss}
\end{equation}
%_____________________________
%_____________________________
with a Curie temperature $\theta\simeq 1.2(2)$\,K.
Measurements performed on a powdered sample (not shown), which are
less affected by anisotropy effects, also follow
Eq.~(\ref{Curie-Weiss}) albeit with a smaller $\theta = 0.5 (1)$\,K
[for the interplay of interactions and anisotropy, see
Eqs.~(3.2)--(3.6) in Ref.~\cite{jongar2001prb}].
These finite $\theta$ point actually to the presence of dipolar
interactions between the molecular spins in the crystal, which would
give rise to long-range order at sufficiently low temperatures.
In the (super)paramagnetic regime of interest here, interactions
merely produce a susceptibility somewhat larger than that of
non-interacting clusters.

In addition to interactions, a most important influence on the
temperature-dependent susceptibilities is exerted by the magnetic
anisotropy \cite{carlin}.
To illustrate these effects, it is helpful to normalize the
experimental $\chi_{T}$ and $\chi_{3 T}$ by their isotropic limits,
$\chi_{\rm iso}$ and $\chi_{3{\rm iso}}$.
These can be obtained from the field expansion coefficients of the
Brillouin magnetization (Appendix~\ref{app:Xiso}), and read
%_____________________________
%
% ISOTROPIC SUSCEPTIBILITIES
%_____________________________
\begin{eqnarray}
\chi_{\rm iso}
&=&
N_{\rm A}
\frac
{(g\mu_{\rm B})^{2}S(S+1)}
{3k_{\rm B} T}
\label{isotropic1}
\\
\chi_{3{\rm iso}}
&=&
-
N_{\rm A}
\frac
{(g\mu_{\rm B})^{4}}
{45(\kT)^3} S(S+1)[S(S+1)+\tfrac{1}{2}]
\;,
\label{isotropic2}
\end{eqnarray}
%_____________________________
%_____________________________
where $N_{\rm A}$ is the number of molecules per mol.
The first of these is merely Curie's law and the second its nonlinear
counterpart.
Normalized by $\chi_{\rm iso}$ and $\chi_{3{\rm iso}}$, the
experimental susceptibilities lose their bare $1/T$ and $1/T^{3}$
contributions, so that their remaining temperature dependence is
mostly due to the effects of the anisotropy.

\begin{figure}
\hspace*{-2.em}
\resizebox{9.cm}{!}{\includegraphics{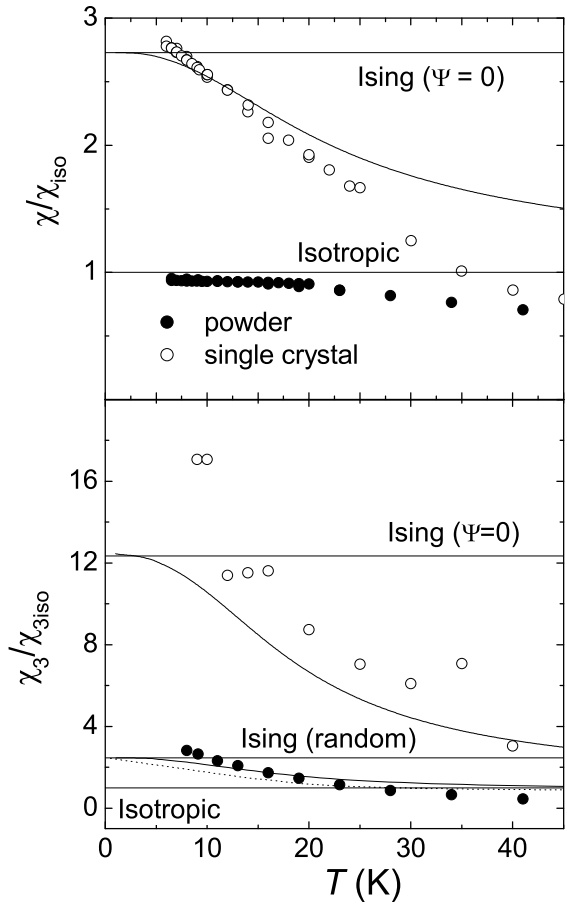}}
\vspace*{-5.ex}
\caption{
Temperature dependence of the equilibrium linear and nonlinear
susceptibilities normalised by their values for isotropic spins
(Brillouin limit).
Upper panel: linear susceptibility; lower panel: nonlinear
susceptibility.
Results are shown for a single crystal of Mn$_{12}$ with the field
parallel to the anisotropy axis $\circ$ and a powder sample $\bullet$.
The lines are theoretical results for classical spins (dotted) and
quantum spins (solid).
}
\vspace*{-3.ex}
\label{XnormvsT}
\end{figure}

The so normalized equilibrium susceptibility data are shown in
Fig.~\ref{XnormvsT}.
Clearly, the isotropic limit is only attained for sufficiently high
temperatures ($T\gtrsim30$\,K).
Note that the high-$T$ limits of $\chi_{T}$ and $\chi_{3T}$ are
slightly smaller than $\chi_{\rm iso}$ and $\chi_{3{\rm iso}}$.
This is caused by the thermal population of higher-energy spin
multiplets of the cluster, the lowest of which has $S=9$ in Mn$_{12}$.
Therefore, in that temperature range the Mn$_{12}$ molecule can no
longer be seen as a superparamagnetic spin $S=10$ and the thermal
mixture of spin states reduces its susceptibilities (the analogue to
the decrease of the spontaneous magnetization in a solid by excitation
of spin-waves).

As the temperature decreases, both $\chi_{T}/\chi_{\rm iso}$ and
$\chi_{3T}/\chi_{3{\rm iso}}$ increase, departing from $\simeq1$.
This is natural since Eqs.~(\ref{isotropic1})--(\ref{isotropic2}) are
only valid when the thermal energy $\kT$ is larger than all zero-field
splittings (produced by the magnetic anisotropy).
The simplest Hamiltonian that describes the magnetic behavior of an
isolated Mn$_{12}$ molecule contains the Zeeman plus uniaxial
anisotropy terms (see also Fig.~\ref{fig:levels})
%_____________________________
%
% SPIN HAMILTONIAN
%_____________________________
\begin{equation}
{\cal H}
=
-DS_{z}^{2}
-
A_{4} S_{z}^{4}
-
g\mu_{\rm B}
(H_{x}S_{x} + H_{y}S_{y} + H_{z}S_{z})
\;.
\label{Hamanis}
\end{equation}
%_____________________________
%_____________________________
Here $D\simeq 0.6$\,K and $A_{4}\simeq 10^{-3}$\,K are the second and
fourth-order anisotropy constants for Mn$_{12}$ \cite{bargatses97},
and $H_{x,y,z}$ the components of the field along the $(a,b,c)$
crystallographic axes.
The largest zero-field splitting produced by the anisotropy occurs
between the states $m=\pm (S-1)$ and the ground state $m=\pm S$:
%_____________________________
%
% ZERO FIELD SPLITTING
%_____________________________
\begin{equation}
\label{W0}
%%%%%%%%%%%%%%%%%%
\Omega_{0}
=
(2S-1)D + [S^{4} -(S-1)^{4}]A_{4}
\simeq
%19\times D + 3439\times A_{4}
%11.4 + 3.4=
14.8
\,{\rm K}
%%%%%%%%%%%%%%%%%%
\;.
\end{equation}
%_____________________________
%_____________________________
When $\kT$ becomes comparable to $\Omega_{0}$ several related effects
occur: (i) the magnetization is no longer given by the Brillouin law
and $T$ does not appear in the combination $H/T$, (ii) $\chi_{T}$ and
$\chi_{3T}$ deviate from the simple
Eqs.~(\ref{isotropic1})--(\ref{isotropic2}) and depend on $\psi$, and
(iii) the normalised susceptibilities adquire a dependence on $T$.
For classical spins, these effects were studied in
Ref.~\cite{gar2000}.
%_____________________________
%_____________________________
%_____________________________
%_____________________________
%_____________________________
%_____________________________

Eventually, in the low temperature limit $\kT/\K\to 0$, only the
states $m=\pm S$ are appreciably populated and each molecular spin
becomes effectively a two-level system (``spin-up'' and ``spin-down''
states).
In this ``Ising'' limit, $\chi_{T}$ and $\chi_{3T}$ can also be
calculated explicitly
%_____________________________
%
% ISING SUSCEPTIBILITIES
%_____________________________
\begin{eqnarray}
\chi_{\rm Ising}
&=&
N_{\rm A}
\frac
{(g\mu_{\rm B})^{2}S^{2}}
{\kT}
\cos^{2}\psi
\label{Ising1}
\\
\chi_{3 {\rm Ising}}
&=&
-N_{\rm A}
\frac{(g\mu_{\rm B})^{4} S^{4}}
{3 (\kT)^{3}}
\cos^{4}\psi
\;.
\label{Ising2}
\end{eqnarray}
%_____________________________
%_____________________________
Here we see that the bare $1/T$ and $1/T^{3}$ dependences are
recovered, so that the normalised susceptibilities become again
temperature independent.
Comparison of these expressions with
Eqs.~(\ref{isotropic1})--(\ref{isotropic2}) reveals that, for a single
crystal at $\psi=0$, $\chi_{T}/\chi_{\rm iso}$ and
$\chi_{3T}/\chi_{3{\rm iso}}$ should increase, respectively, by an
overall factor of
$2.7$ [$=3\times S^{2}/S(S+1)$] and
$12.3$ ($=15\times S^{4}/S(S+1)[S(S+1)+\tfrac{1}{2}]$),
when decreasing temperature.
The experimental curves approach indeed these values at low
temperatures ($T<10$\,K in Fig.~\ref{XnormvsT}) although they become
even larger.
This is probably due to the interactions which, as we have seen,
enhance the magnetic response at low temperatures.

Anyhow, these results suggest the measurement of the
temperature-dependent reduced susceptibilities as a suitable tool to
estimate the anisotropy parameters of superparamagnets.
In this respect, the advantage of the nonlinear susceptibility becomes
evident when dealing with systems with randomly oriented axes.
Then the ratios between the ``Ising'' and isotropic limits decrease
significantly [cf.\ Eqs.~(\ref{isotropic1})--(\ref{isotropic2}) with
Eqs.~(\ref{Ising1})--(\ref{Ising2})].
Indeed, $\chi_{T}/\chi_{\rm iso}$ becomes nearly $T$-independent,
whereas the reduced nonlinear susceptibility still retains a sizable
variation with $T$ (by a factor $\sim2.5$).
This is experimentally confirmed by measurements on a polycrystalline
sample (Fig.~\ref{XnormvsT}) \cite{endnote43}.
Thus we see that, contrary to the linear response, $\chi_{3T}$ keeps
information on the anisotropy even for superparamagnets with axes
distributed at random.
This is the case most often encountered in nanoparticle systems
\cite{panpol93} but also for single-molecule magnets when deposited on
surfaces \cite{surfaces} or inside porous materials \cite{porous}.

The considerations above can be supported by direct diagonalization of
the Hamiltonian~(\ref{Hamanis}).
The results (solid lines in Fig.~\ref{XnormvsT}) exhibit the same
trends as the experiments, both for parallel axes and after averaging
over random orientations.
Full agreement is precluded by the effect of interactions, on the
low-$T$ side, and by the population of excited multiplets with $S\neq
10$ at high $T$, as discussed above.

Before concluding this section, there is an additional feature that
deserves to be commented upon.
Consider the theoretical behavior of $\chi_{3T}$ in {\em classical}
spins, also plotted in Fig.~\ref{XnormvsT} \cite{endnote44}.
We see that, although classical and quantum calculations predict the
crossover from the isotropic to the Ising limits, the classical
susceptibilities are shifted towards lower temperatures.
This shift can be seen as a manifestation of the quantum, discrete
nature of the energy spectrum of Mn$_{12}$.
The finite energy gap between the two lowest quantum levels,
$\Omega_{0}$, leads to a faster (exponential in $\K/T$) convergence to
the Ising limit, whereas classically this limit is only approached
with a slow power law in $\K/T$ (see Appendix~\ref{app:Xani}).

%%%%%%%%%%%%%%%%%%%%%%%%%%%%%%%%%%%%%%%%%%%%%%%%%%%%%%%%%%%%
%%%%%%%%%%%%%%%%%%   DYNAMICAL  RESPONSE   %%%%%%%%%%%%%%%%%
%%%%%%%%%%%%%%%%%%   DYNAMICAL  RESPONSE   %%%%%%%%%%%%%%%%%
%%%%%%%%%%%%%%%%%%   DYNAMICAL  RESPONSE   %%%%%%%%%%%%%%%%%
%%%%%%%%%%%%%%%%%%   DYNAMICAL  RESPONSE   %%%%%%%%%%%%%%%%%
%%%%%%%%%%%%%%%%%%   DYNAMICAL  RESPONSE   %%%%%%%%%%%%%%%%%
%%%%%%%%%%%%%%%%%%%%%%%%%%%%%%%%%%%%%%%%%%%%%%%%%%%%%%%%%%%%

\section{
Dynamical susceptibilities
}
\label{dynamical}

In this section we turn our attention from the equilibrium to the
dynamical response.
We begin reviewing briefly the behavior of the nonlinear
susceptibility $\Xnl$ in the {\em classical\/} case.
This allows to introduce some basic expressions valid also for quantum
superparamagnets; then we present the experimental results for
Mn$_{12}$.

%%%%%%%%%%%%%%%%%%%%%%%%%%%%%%%%%%%%%%%%%%%%%%%%%%%%%%%%%%%%
%%%%%%%%%%%%%%%%%%%%%   CLASSICAL X3   %%%%%%%%%%%%%%%%%%%%%
%%%%%%%%%%%%%%%%%%%%%   CLASSICAL X3   %%%%%%%%%%%%%%%%%%%%%
%%%%%%%%%%%%%%%%%%%%%   CLASSICAL X3   %%%%%%%%%%%%%%%%%%%%%
%%%%%%%%%%%%%%%%%%%%%%%%%%%%%%%%%%%%%%%%%%%%%%%%%%%%%%%%%%%%

\subsection{
Classical superparamagnets and modelization
}

The dynamical nonlinear susceptibility of classical spins was
theoretically found to be very large and, in contrast to the linear
susceptibility, non-trivially sensitive to the spin-bath coupling
strength $\lambda$ \cite{garsve2000,gargar2004}.
The ``damping'' $\lambda$ measures the relative importance of the
relaxation and Hamiltonian (precessional) terms in the dynamical
equations \cite{bro63,gar2000}.
Thus $1/\lambda$ is of the order of the number of precessional turns
that the spin executes in the deterministic spiraling down to the
energy minima.

The contributions to the nonlinear response of the longitudinal and
transverse components of the field are captured by a simple formula
involving the low-$H$ expansion coefficients $\glo$ and $\gtr$ of the
relaxation rate \cite{jongar2001epl}
%_____________________________
%
% EXPANSION RELAXATION RATE
%_____________________________
\begin{equation}
\rate
\simeq
\rateo
\big[
1
+
\tfrac{1}{2}
\big(
\glo
\xi_{\|}^{2}
+
\gtr
\xi_{\perp}^{2}
\big)
\big]
\;,
\label{expansionG}
\end{equation}
%_____________________________
%_____________________________
where $\rateo=\rate|_{H=0}$ and $\xi=g\mu_{\rm B}SH/\kT$
\cite{endnote45}.
The expression for the nonlinear susceptibility oscillating with the
third harmonic of the field \cite{gargar2004} can be found in
Appendix~\ref{app:X3}.
It is easy to find the counterpart for the $\Xnl$ oscillating with
${\rm e}^{\pm\iu\,\w t}$, as used in this work, which reads
%_____________________________
%
% NONLINEAR SUSCEPTIBILITY + RATE EXPANSION
% ISING
%_____________________________
\begin{eqnarray}
\Xnl
&=&
-N_{\rm A}
\frac{(g\mu_{\rm B})^{4} S^{4}}
{3 (\kT)^{3}}
\bigg[
\frac{\cos^{4}\psi}{1+\iu\,\w\tau}
\nonumber\\
& &
{}-
\frac{\iu\,\w\tau}{2( 1+\iu\,\w\tau)^{2}}
(\glo\cos^{4}\psi+ \gtr\cos^{2}\psi\sin^{2}\psi)
\bigg]
\qquad
\label{X3:debye}
\end{eqnarray}
%_____________________________
%_____________________________
Here the Ising approximation for the equilibrium parts has been used
(this works fine at temperatures around the blocking temperature
$T_{\rm b}\sim 5$\,K; see Sec.~\ref{sec:equilibrium}).
The longitudinal part, proportional to $\cos^{4}\psi$, is maximum at
$\psi=0$ (in absolute value); the ``transverse'' contribution
associated to $\gtr$ becomes zero both at $\psi=0$ and $\pi/2$, being
maximum at $\psi=\pi/4$.
Equation~(\ref{X3:debye}) shows that the magnitude, signs, and the
$\w$ and angular dependences of $\Xnl$ are determined by the
competition between the rate expansion coefficients $\glo$ and $\gtr$.
Therefore, measurements of those dependences can provide valuable
information on the different contributions to the spin reversal.
%_____________________________
%_____________________________
%_____________________________
\begin{figure}
\hspace*{-2.em}
\resizebox{9.cm}{!}{\includegraphics{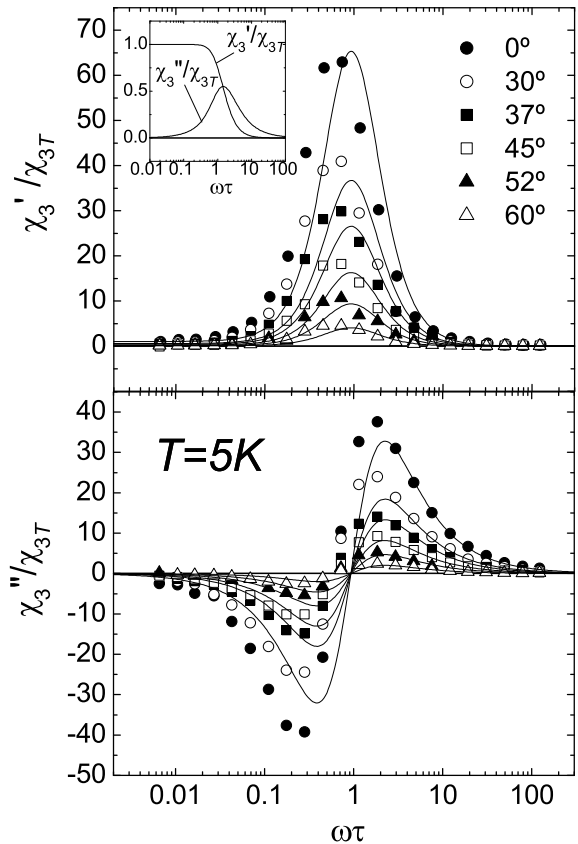}}
\vspace*{-5.ex}
\caption{
Nonlinear susceptibility vs.\ frequency at $T=5$\,K.
Results are shown for different angles $\psi$ between the applied
field and the anisotropy axis.
The data are normalized by the equilibrium $\chi_{3T}$ measured at
$\psi=0$.
The relaxation time $\tau$ was obtained from $\w$-dependent linear
susceptibility measurements (as those of Fig.~\ref{X1vsfreq}).
Upper panel: real part $\Xnl'$; lower panel: imaginary part $\Xnl''$.
The lines are obtained using Eq.~(\ref{X3:debye}) with $\glo=-260$
and $\gtr=0$.
The inset shows the sharply different classical prediction for
$\psi=0$.
}
\vspace*{-3.ex}
\label{X3normvsfreq}
\end{figure}
%_____________________________
%_____________________________
%_____________________________

For classical superparamagnets, where only the thermoactivation
operates, the rate expansion coefficients are given in the considered
low-temperature range by \cite{gargar2004,jongar2001epl}
%_____________________________
%_____________________________
\begin{equation}
\label{Gamma:coeffs:app}
\glo
=
1
\;,
\qquad
\gtr
=
F(\lambda)/2
\;,
\end{equation}
%_____________________________
%_____________________________
where $F>0$ is a function of $\lambda$ (and $T$).
For strong damping, $F\to1$, so that $\glo$ and $\gtr$ are of the same
order.
In contrast, in the weak damping regime (governed by the precession),
one has $F\propto1/\lambda$ and $\gtr$ becomes very large, dominating
the nonlinear response.
Then one can find that the real part $|\Xnl'|\gg|\chi_{3 T}|$ but
with $\Xnl'/\chi_{3 T}<0$ (i.e., the sign is reversed with respect
to the equilibrium value).
This phenomenology is equivalent to that of the third harmonic
nonlinear susceptibility \cite{garsve2000,gargar2004}, since both
quantities exhibit similar dependences on $\lambda$ and $\w$.

It is worth mentioning that in the derivation of Eq.~(\ref{X3:debye})
rather general assumptions are invoked (Appendix~\ref{app:X3}).
Therefore, the functional form obtained is quite generic and valid
for classical as well as quantum superparamagnets.
Then, the relevant information on the quantum reversal mechanisms will
be incorporated by the rate expansion coefficients $\glo$ and $\gtr$.
Naturally, they could be very different from their classical
counterparts (\ref{Gamma:coeffs:app}).

%%%%%%%%%%%%%%%%%%%%%%%%%%%%%%%%%%%%%%%%%%%%%%%%%%%%%%%%%%%%
%%%%%%%%%%%%%%%%%%%%%   X3  FOR  Mn12   %%%%%%%%%%%%%%%%%%%%
%%%%%%%%%%%%%%%%%%%%%   X3  FOR  Mn12   %%%%%%%%%%%%%%%%%%%%
%%%%%%%%%%%%%%%%%%%%%   X3  FOR  Mn12   %%%%%%%%%%%%%%%%%%%%
%%%%%%%%%%%%%%%%%%%%%%%%%%%%%%%%%%%%%%%%%%%%%%%%%%%%%%%%%%%%

\subsection{
Nonlinear dynamical susceptibility of Mn$_{12}$
}

After these theoretical considerations let us turn to the experiments
on quantum nanomagnets.
Figure~\ref{X3normvsfreq} displays frequency-dependent measurements of
the nonlinear susceptibility of Mn$_{12}$ performed at constant $T$
for different angles $\psi$.
They demonstrate that the result already shown in Fig.~\ref{XvsT},
namely, that $\Xnl'$ becomes much larger than its equilibrium value
$\chi_{3T}$ near the blocking temperature, is a dynamical effect not
caused by magnetic ordering or some kind of ``freezing'' (interactions
also enhance the susceptibilities, but by a much smaller factor).
We know that classically one can also have $|\Xnl'|\gg|\chi_{3
T}|$, but here $\Xnl'/\chi_{3 T}>0$, that is, the peak of the
measured nonlinear dynamic susceptibility is {\em reversed\/} with
respect to the classical prediction.

From Eq.~(\ref{X3:debye}) we see that at $\psi=0$ there is no
contribution of $\gtr$ to $\Xnl$.
In addition, the first term in that equation, which has a Debye-type
profile, cannot provide $|\Xnl'|>|\chi_{3 T}|$ because ${\rm
Re}[\chi/(1+\iu x)]\leq\chi$.
Therefore, the maximum observed in Fig.~\ref{X3normvsfreq} should be
due to the $\glo$ contribution.
There is a result relating the height of the susceptibility peak
$\Xnl'|_{\rm max}$ with the combination
$\QFt(\psi)\equiv\glo\cos^{2}\psi+\gtr\sin^{2}\psi$ of the relaxation
rate coefficients (Appendix~\ref{app:X3})
%_____________________________
%
% MAXIMUM REAL X3
%_____________________________
\begin{equation}
\label{maximum:Re:height:2}
\Xnl'|_{\rm max}/\chi_{3T}\big|_{\psi=0}
\simeq
-\cos^{2}\psi\,\QFt(\psi)/4
\;.
\end{equation}
%_____________________________
%_____________________________
Therefore, the positive sign of the maximum of
$\Xnl'(\w)/\chi_{3T}$ at $\psi=0$ entails $\QFt<0$.
But $\QFt|_{\psi=0}=\glo$, entailing that the relaxation time
$\tau=1/\rate$ becomes {\em longer\/} as $\Hlo$ increases.
No classical mechanism can account for this; actually, $\glo=1$ in the
classical model, which gives a $\Xnl'/\chi_{3T}|_{\psi=0}$
smaller than $1$ and decreasing with $\w$ (inset of
Fig.~\ref{X3normvsfreq}), in sharp contrast to the measured $\Xnl$.
On the other hand, it is well-established
\cite{frietal96,heratal96,thoetal96,garchu97,luibarfer98,wur98} that
in Mn$_{12}$ the suppression of tunneling by a longitudinal field {\em
strongly\/} reduces the relaxation rate (as it breaks the degeneracy
between initial ``spin-up'' and final ``spin-down'' states, inhibiting
the tunnel channels).
As suggested in Ref.~\cite{luietal2004a} this effect provides the
$\glo$ required, both negative and large, to account for the
experimental $\Xnl$ of Mn$_{12}$.
Thus, we see that the known field-suppression of tunneling shows up in
the nonlinear response as a distinctly quantum contribution, with its
sign reversed with respect to the classical case.
%_____________________________
%_____________________________
%_____________________________
\begin{figure}
\hspace*{-2.em}
\resizebox{9.cm}{!}{\includegraphics{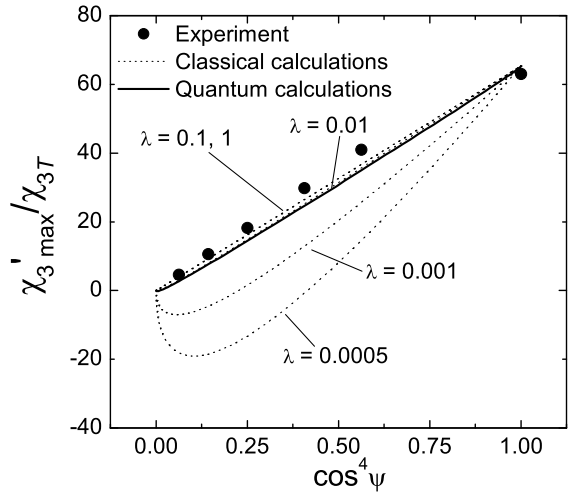}}
\vspace*{-5.ex}
\caption{
Evolution of the maximum of $\Xnl'(\w)$ with the angle $\psi$ of
the applied field at $T=5$\,K.
The symbols are the experimental results.
The dashed lines are obtained from Eq.~(\ref{X3:debye}) using the
experimentally determined $\glo=-260$ and the classical transverse
contribution $\gtr=F(\lambda)/2$ computed for different values of the
phenomenological damping constant $\lambda$.
The solid line corresponds to theoretical calculations with a Pauli
quantum master equation (see the text).
}
\vspace*{-3.ex}
\label{X3normvspsi}
\end{figure}
%_____________________________
%_____________________________
%_____________________________

When $\psi>0$, the detuning quantum contribution coexists with the
transverse $\gtr$ contribution.
Still, the data of Fig.~\ref{X3normvsfreq} suggest that
$\Xnl\propto\cos^{4}\psi$ holds approximately.
We can check this by accounting again for
Eq.~(\ref{maximum:Re:height:2}) and plotting the maximum of $\Xnl'$
vs.\ $\cos^{4}\psi$ (Fig.~\ref{X3normvspsi}).
This yields an almost straight line indicating that in Mn$_{12}$ the
coefficient $\glo$ overwhelmingly dominates $\gtr$, which in the
classical case embodied the precessional contribution and could be
sizable (we return to this issue in Sec.~\ref{decoh-vs-diss}).

\begin{figure}
\hspace*{-2.em}
\resizebox{9.cm}{!}{\includegraphics{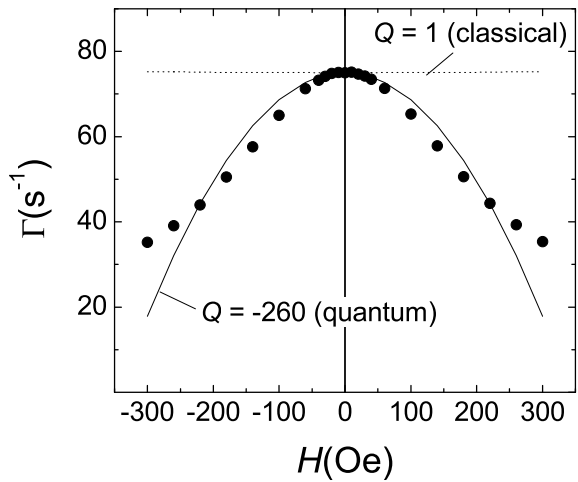}}
\vspace*{-5.ex}
\caption{
Relaxation rate of Mn$_{12}$ at $T=5$\,K as a function of the field
applied along the anisotropy axis ($\psi=0$).
The symbols are the $\rate$'s obtained from Debye fits of
$\chi(\w,H)$.
The lines are calculated as $\rate=\rateo (1+\QFt\,\xi^{2}/2)$, where
$\xi=g\mu_{\rm B} S \Hlo/\kT$ and $\QFt$ comes from the maximum of
$\Xnl'(\w)$ via Eq.~(\ref{maximum:Re:height:2}).
The classical prediction ($\QFt=1$) is shown for comparison.
}
\vspace*{-3.ex}
\label{GammavsH}
\end{figure}

Thus $\Xnl(\w,\psi)$ provides {\em direct\/} experimental access to
the relaxation-rate expansion coefficients $\glo$ and $\gtr$, which
contain information on the spin relaxation mechanisms.
Besides, from the sign of the $\Xnl(\w)$ peaks we can infer
whether the spin reversal is dominated by a classical mechanism or by
quantum processes.
The consistency of our analysis can be ascertained by comparing
the experimentally determined rate $\rate$ (obtained from Debye
fits of $\chi$), with the rate reconstructed from the expansion
$\rate=\rateo(1+\QFt\,\xi^{2}/2+\dots)$, using the
$\QFt|_{\psi=0}$ extracted from the $\Xnl'$ maxima via
Eq.~(\ref{maximum:Re:height:2}) ($\QFt_{{\rm
Mn}_{12}}|_{\psi=0}\simeq-260$).
Figure~\ref{GammavsH} shows the good agreement between both
results in the weak field regime, supporting our interpretation.
%_____________________________
%_____________________________
%_____________________________
%_____________________________
%_____________________________
%_____________________________

%%%%%%%%%%%%%%%%%%%%%%%%%%%%%%%%%%%%%%%%%%%%%%%%%%%%%%%%%%%%
%%%%%%%%%%%%%   DECOHERENCE  VS  DISSIPATION   %%%%%%%%%%%%%
%%%%%%%%%%%%%   DECOHERENCE  VS  DISSIPATION   %%%%%%%%%%%%%
%%%%%%%%%%%%%   DECOHERENCE  VS  DISSIPATION   %%%%%%%%%%%%%
%%%%%%%%%%%%%   DECOHERENCE  VS  DISSIPATION   %%%%%%%%%%%%%
%%%%%%%%%%%%%   DECOHERENCE  VS  DISSIPATION   %%%%%%%%%%%%%
%%%%%%%%%%%%%%%%%%%%%%%%%%%%%%%%%%%%%%%%%%%%%%%%%%%%%%%%%%%%

\section{
Dissipation vs.\ decoherence  in single-molecule magnets
}
\label{decoh-vs-diss}

We have seen in the previous section that the transverse contribution
to the nonlinear susceptibility is nearly absent in Mn$_{12}$ (indeed
the experiments are consistent with $\gtr=0$).
Classically, $\gtr$ incorporates the precessional contribution, which
can be large for weak damping $\lambda\ll1$
\cite{garsve2000,gargar2004}.
Still, as its long $\tau_{0}$ indicates, Mn$_{12}$ is expected to be
quite ``underdamped'' (in the sense of energy dissipation).
Thus, it seems that some other process makes the spin to loss its
intrinsic, or coherent, precessional dynamics and appear, when seen
through $\Xnl(\w)$, as ``overdamped''.
In this last section we will try to reconcile these results
($\gtr\simeq0$ and long $\tau_{0}$) invoking an effect of the coupling
to the bath absent in classical physics --- {\em decoherence}.

%%%%%%%%%%%%%%%%%%%%%%%%%%%%%%%%%%%%%%%%%%%%%%%%%%%%%%%%%%%%
%%%%%%%%%%%%%%%   BOUND  FOR  THE  DAMPING   %%%%%%%%%%%%%%%
%%%%%%%%%%%%%%%   BOUND  FOR  THE  DAMPING   %%%%%%%%%%%%%%%
%%%%%%%%%%%%%%%   BOUND  FOR  THE  DAMPING   %%%%%%%%%%%%%%%
%%%%%%%%%%%%%%%%%%%%%%%%%%%%%%%%%%%%%%%%%%%%%%%%%%%%%%%%%%%%

\subsection{
Experimental bound for the effective damping
}

Before starting let us quantify a lower bound for an effective
$\lambda$ of Mn$_{12}$.
To this end we generate $\Xnl(\w)$ curves using Eq.~(\ref{X3:debye})
with the $\glo$ experimentally determined from $\Xnl(\w)|_{\psi=0}$,
while assuming $\gtr=F(\lambda)/2$ as in classical superparamagnets.
In this way we compute $\Xnl'|_{\rm max}$ vs.\ $\cos^{4}\psi$ for
several $\lambda$ and compare with the experimental results
(Fig.~\ref{X3normvspsi}).
The best agreement is obtained for large $\lambda$, which in fact
yields small $\gtr$ and hence almost no $\cos^{2}\psi\sin^{2}\psi$
contribution.
However, taking into account experimental uncertainties as well as the
smaller sensitivity of $\gtr$ to large $\lambda$, gives a lower bound
of $\lambda\gtrsim0.01$.

In the classical equations of motion the damping parameter $\lambda$
is a measure of the relative weights of their relaxation and
precessional terms.
It can then be expressed as the Larmor precession period $\tau_{\rm
L}=2\pi/\gamma H_{\rm a}$ (in the anisotropy field $H_{\rm a}$;
$\gamma=g\mu_{\rm B}/\hbar$) divided by some timescale of relaxation.
For the latter we can use the classical prefactor in the Arrhenius law
\cite{bro63,gar2000}
%_____________________________
%
% PREFACTOR ARRHENIUS LAW: CLASSICAL
%_____________________________
\begin{equation}
\label{tau}
\tau_{0}
=
\frac{1}{\lambda\gamma H_{\rm a}}
\sqrt{\frac{\pi}{4\,\sigma}}
\,
\Big(
1
+
\frac{1}{\sigma}
%
%+\frac{7}{4\sigma^{2}}
+
\cdots
\Big)
\;,
\end{equation}
%_____________________________
%_____________________________
where the reduced anisotropy barrier is $\sigma=U_{0}/\kT$.
In our experiments at $T=5$\,K we have $\sigma\simeq14$, whence
%_____________________________
%
% DAMPING COEFF. & TIME SCALES
%_____________________________
\begin{equation}
\label{lambda}
\lambda
\simeq
0.04
\times
\tau_{\rm L}/\tau_{0}
\;.
\end{equation}
%_____________________________
%_____________________________
Experimentally $\tau_{0}\sim 3\times 10^{-8}$\,s in Mn$_{12}$.
Its anisotropy field can be obtained from magnetization measurements
(along the hard plane) or from the ground-state transition frequency
$\Omega_{0}/\hbar$ [Eq.~(\ref{W0})], getting $\tau_{\rm
L}\simeq3$--$4\times 10^{-12}$\,s.
This gives $\tau_{\rm L}/\tau_{0}\sim10^{-4}$ and hence an effective
$\lambda$ many orders of magnitude smaller than the lower bound
$\lambda\gtrsim 0.01$ extracted from $\Xnl$.

The estimation of $\tau_{\rm L}/\tau_{0}$ is in agreement with the
mentioned underdamped character of Mn$_{12}$ (in the sense of
long-lived levels).
However, in case of $\gtr$ having some precession-type contribution
akin to $F(\lambda)$, the above $\lambda\propto\tau_{\rm L}/\tau_{0}$
would not be the parameter entering there.
Otherwise a giant $\gtr\propto1/\lambda$ would dominate $\Xnl$
(leading to $\Xnl'/\chi_{3 T}<0$ and
$\Xnl'|_{\rm max}$ not proportional to $\cos^{4}\psi$),
which is clearly not seen in the experiments.
Still, it might be that, instead of the damping, some other timescale
limits the precession; as this is a type of coherent dynamics, we can
call such time a decoherence time $\tdec$.
This should replace $\tau_{0}$ in the effective
$\lambda\propto\tau_{\rm L}/\tdec$.
To get a $\lambda$ consistent with the lower bound
$\lambda\gtrsim0.01$ obtained from $\Xnl$, the time $\tdec$ should
be far shorter than $\tau_{0}\sim 10^{-8}$\,s, actually
$\tdec\sim1$--$1.5\times 10^{-11}$\,s.

%%%%%%%%%%%%%%%%%%%%%%%%%%%%%%%%%%%%%%%%%%%%%%%%%%%%%%%%%%%%
%%%%%%%%%%%%%%%%%%   QUANTUM  TREATMENT   %%%%%%%%%%%%%%%%%%
%%%%%%%%%%%%%%%%%%   QUANTUM  TREATMENT   %%%%%%%%%%%%%%%%%%
%%%%%%%%%%%%%%%%%%   QUANTUM  TREATMENT   %%%%%%%%%%%%%%%%%%
%%%%%%%%%%%%%%%%%%%%%%%%%%%%%%%%%%%%%%%%%%%%%%%%%%%%%%%%%%%%

\subsection{
Approximate quantum treatment
}

A critical assessment of the considerations above is in order
before proceeding.
They have been based on stretching the classical idea that $\gtr$
should include some precessional-type contribution.
Further, they assume that this contribution is controlled by a
parameter relating the Larmor period with some scale limiting the time
allowed to the spin to precess (either by damping or by some loss of
coherence).
By definition, however, $\gtr$ accounts for the effects of $\Htr$ on
the relaxation rate [Eq.~(\ref{expansionG})]; besides, $\Xnl(\w)$
gives {\em direct\/} access to $\glo$ and $\gtr$.
Then, one would expect that a pure quantum approach for $\rate$,
including the bath coupling and coherent dynamics (tunnel and
precession) could account for the experimental $\Xnl$ without
recourse to classically preconceived notions.

An exact quantum treatment, unfortunately, is difficult because
one must deal with the full density-matrix equation including the
intrinsic (Hamiltonian) dynamics plus the effects of the bath
(damping and decoherence).
However, as this can be handled in various limit cases, we shall
attempt a discussion based on the corresponding partial solutions
for the quantum $\rate$, with the hope of shedding some light on
the physical origin of the results.

The dominant terms that enter the relaxation rate describing quantum
tunneling via a pair of nearly degenerate states $|{\pm m}\rangle$ has
a Lorentzian shape as a function of the {\em longitudinal\/} bias
$\xi_{m}=2 m g\mu_{\rm B}H_{\|}$ \cite{frisarzio98}
%_____________________________
%
% QUANTUM RATE
%_____________________________
\begin{equation}
\rate
\simeq
2\Gamma_{m}
\frac
{\Delm^{2}}
{\xi_{m}^{2}
+
\underbrace{\Delm^{2}+\hbar^{2}\Gamma_{m}^{2}}_{w_{m}^{2}}
}
\exp (-U_{m}/\kT)
\;.
\label{GammaQ}
\end{equation}
%_____________________________
%_____________________________
Here $\Gamma_{m}$ is the probability of decaying to other levels via
the absorption or emission of phonons (i.e., it is approximately
$1/\tau_{0}$), $\Delm$ is the tunnel splitting of the pair
$|{\pm m}\rangle$ induced by terms not commuting with $S_{z}$ in the
Hamiltonian, and $U_{m}$ is the energy of the levels.
The width of the Lorentzian introduced \cite{tupbar2002},
$w_{m}^{2}=\Delm^{2}+\hbar^{2}\Gamma_{m}^{2}$, interpolates between
the results that can be obtained in the two limit cases of (i) large
coupling $\hbar/\tau_{0}\gg\Delm$ \cite{garchu97}, where
$w_{m}\simeq\hbar/\tau_{0}$, and (ii) weak coupling
$\hbar/\tau_{0}\ll\Delm$ \cite{luibarfer98}, in which
$w_{m}\simeq\Delm$.

Performing the second $\Htr$-derivative of Eq.~(\ref{GammaQ}) gives
the corresponding $\gtr$.
For the parameters of Mn$_{12}$ it results that the main contribution
at zero longitudinal bias comes from the derivative of the quotient
$\Delm^{2}/w_{m}^{2}$
%_____________________________
%
% G PERP: QUANTUM
%_____________________________
\begin{equation}
\gtr
\equiv
\frac{1}{\rateo}\frac{\partial^{2}
\rate}{\partial \Htr^{2}}
\Big|_{0}
\simeq
\frac{2}{(\Delm/w_{m})}
\frac
{\partial^{2}}
{\partial \Htr^{2}}
(\Delm/w_{m})
\;.
\label{GperpQ}
\end{equation}
%_____________________________
%_____________________________
Now, for large coupling $\hbar/\tau_{0}\gg\Delm$, we have
$w_{m}=\hbar/\tau_{0}$.
Then the relaxation rate depends on the ratio between $\tau_{0}$ and
the tunneling time $\hbar/\Delm$.
A transverse field eases the tunneling and significantly increases
$\Delm$, and hence $\rate$.
As $\rate$ is then quite sensitive to $\Htr$, we can have large
$\gtr$, in analogy with the classical situation.
On the contrary, when $\hbar/\tau_{0}\ll\Delm$, we have
$w_{m}\simeq\Delm$ and hence $\Delm^{2}/w_{m}^{2}\simeq1$.
Then the rate becomes quite insensitive to $\Delm$ (and hence to
$\Htr$), leading to a small $\gtr$.

In Mn$_{12}$, where $\hbar/\tau_{0}\sim 0.2$\,mK, both situations are
possible.
The reason is the exponential increase of the tunnel splitting $\Delm$
with decreasing $|m|$, going from the sub-nanoKelvin regime for the
ground levels $m=\pm S$ to some tenths of K for $|m|\lesssim 2$.
Therefore, the relation between $\hbar/\tau_{0}$ and $\Delm$ depends
on which tunneling path (i.e., which pair $\pm m$) gives the dominant
contribution to $\rate$.
If tunneling proceeds via the deep levels, where
$\hbar/\tau_{0}\gg\Delm$, we would find a large $\gtr$.
In contrast, when tunneling occurs through the excited levels, one has
$\hbar/\tau_{0}\ll\Delm$ and hence small $\gtr$.

At this point it is important to bring into the discussion the effect
of environmental bias fields (due to intermolecular dipolar
interactions or the hyperfine interaction with the nuclear spins of
the Mn ions).
They produce a distribution of bias $\xi$ whose typical width is of a
few tenths of K (of the order of the Curie-Weiss $\Theta$).
These bias fields enter as $\Delm^{2}/(\xi_{m}^{2}+w_{m}^{2})$ in the
rate expression (\ref{GammaQ}), replacing the bare
$\Delm^{2}/w_{m}^{2}$ and supressing tunneling when $w_{m}\ll\xi_{m}$
(Fig.~\ref{fig:levels}).
Taking into account the order of magnitude of $\xi_{m}$, the bias
effectively blocks tunneling via the large $|m|$ (deep) channels,
those that would provide large $\gtr$.
Tunneling becomes possible only for $\Delm>\xi_{m}$, but for those
upper levels $\hbar/\tau_{0}\ll\Delm$, leading to small $\gtr$, in
agreement with the experiments.

We can support this picture with direct numerical calculations.
An approximate (Pauli) quantum master equation, which works well when
tunneling occurs under weak damping conditions and that incorporates
the effects of environmental bias fields, was used to study several
problems in Mn$_{12}$ \cite{luibarfer98,luimetjon2002,ferluibar98jap}.
We have implemented it to address the nonlinear response problem,
mimicking the experimental protocol and calculating $\chi'$ and
$\chi''$ vs.\ $H$.
Results are shown in Fig.~\ref{X3normvspsi} (solid line; see also
Fig.~3 of Ref.~\cite{luietal2004a}).
They account well for the measured nonlinear susceptibility; in
particular, for the nearly $\cos^{4}\psi$ dependence of $\Xnl$
associated to small $\gtr$.

We would like to also provide a physical picture in the limit cases
discussed ($\hbar/\tau_{0}$ much larger or smaller than $\Delm$).
To this end, let us discuss the total energy of the spin plus the
bath, treating their interaction perturbatively.
Then, time-dependent perturbation theory leads to the celebrated {\em
  time-energy\/} ``uncertainty'' relation.
In particular, for $t\gtrsim\hbar/\Delta E$ the dominant transitions
are those conserving the total energy.
On the other hand, for a spin in a $|m\rangle$ state, which is not an
exact eigenstate of the Hamiltonian, the energy uncertainty is of the
order of the tunnel splitting $\Delm$.
Well, consider now a transition to such an $|m\rangle$ state; although
the spin could at short times remain there, for times longer than
$\tphi\equiv\hbar/\Delm$ it will have to become an energy eigenstate.
Then the wavefunction consists of a superposition of spin-up and
spin-down states $|{\pm m}\rangle$, delocalized between both sides of
the barrier.

We can now revisit the limit cases discussed above.
Consider first the strong-coupling case $\hbar/\tau_{0}\gg\Delm$.
For times shorter than the decay $\tau_{0}$, the time-energy
uncertainty allows the existence of superpositions of energy
eigenstates, which can be localized on either side of the barrier
($\sim|{\pm m}\rangle$).
These wavepackets may exhibit Hamiltonian dynamics including tunnel
oscillations and precession.
Under these conditions the rate $\rate$ is quite sensitive to
$\Delm$, as it controls the probability for the spin to have
tunneled before a time $\tau_{0}$.
It is this sensitivity to $\Delm$, and in turn to $\Htr$, which can
lead to large $\gtr$.

When, by constrast, $\hbar/\tau_{0}\ll\Delm$ the ``semiclassical''
wavepacket around $|m\rangle$ delocalizes in the tunneling time
$\tphi=\hbar/\Delm$, evolving towards an energy eigenstate due to the
uncertainty principle.
Then, there is no coherent oscillation between $|m\rangle$ and
$|{-m}\rangle$, neither precession of the (averaged) transverse
components, since this is a stationary state.
The dependence on $\Delm$ (and hence on $\Htr$) is then minimized, as
the wavefunction is already delocalized between the spin-up and
spin-down states, leading to small $\gtr$ values.
Note that under these conditions the coherent (precessional) dynamics
is not limited by the level lifetime $\tau_{0}$ but by the much
shorter ``decoherence'' time to attain a diagonal density matrix.
Therefore, the $\tdec$ introduced heuristically above can be
identified with this $\tphi=\hbar/\Delm$.
For $|m|=2$, $4$, we have $\Delm\sim0.7$--$0.02$\,K, which give
$\tphi\sim10^{-11}$--$4\times 10^{-10}$\,s.
These values are consistent with the estimation, based on $\Xnl(\w)$,
of the $\tdec$ required to get $\lambda\sim0.01$ (which yielded
$\tdec\sim10^{-11}$\,s) \cite{endnote46}.

%%%%%%%%%%%%%%%%%%%%%%%%%%%%%%%%%%%%%%%%%%%%%%%%%%%%%%%%%%%%
%%%%%%%%%%%%%   DISCUSSION  AND  CONCLUSIONS   %%%%%%%%%%%%%
%%%%%%%%%%%%%   DISCUSSION  AND  CONCLUSIONS   %%%%%%%%%%%%%
%%%%%%%%%%%%%   DISCUSSION  AND  CONCLUSIONS   %%%%%%%%%%%%%
%%%%%%%%%%%%%   DISCUSSION  AND  CONCLUSIONS   %%%%%%%%%%%%%
%%%%%%%%%%%%%   DISCUSSION  AND  CONCLUSIONS   %%%%%%%%%%%%%
%%%%%%%%%%%%%%%%%%%%%%%%%%%%%%%%%%%%%%%%%%%%%%%%%%%%%%%%%%%%

\section{
Summary and conclusions
}
\label{conclusions}

The single-molecule magnet Mn$_{12}$ is a model system for the study
of thermal-equilibrium properties of spins with magnetic anisotropy,
as well as the dynamics of a quantum mesoscopic system subjected to
the effects of a dissipative environment.
In this system, to the rich physics of classical supeparamagnets,
quantum effects are incorporated.
In this article we have investigated experimentally the equilibrium
and dynamical nonlinear responses of Mn$_{12}$.
The nonlinear susceptibility $\Xnl$ was underexploited in this field
in spite of having, when compared to the linear susceptibility, an
enhanced sensitivity to several important characteristics of the
system.

We have shown the sensitivity of the {\em equilibrium\/} $\Xnl$ to the
magnetic anisotropy parameters of the spin Hamiltonian.
As in the classical case, the anisotropy leads to an extra temperature
dependence of $\Xnl$ that, in contrast to the linear susceptibility,
persists for randomly distributed anisotropy axes.
Therefore, the measurement of $\Xnl(T)$ can be exploited to estimate
the anisotropy constants even in powdered samples or in systems
deposited on surfaces or in porous materials.

The analysis of the {\em dynamical\/} $\Xnl$, with help from a generic
but simple formula [Eq.~(\ref{X3:debye})], provides valuable
information on the intrinsic dynamics of the system.
Specifically, $\Xnl(\w,\psi)$ gives access to the relaxation-rate
field-expansion coefficients $\glo$ and $\gtr$
[Eq.~(\ref{expansionG})], which contain information on the mechanisms
of spin reversal.
The experimental nonlinear response of Mn$_{12}$ is found to be
consistent with the established scenario of thermally activated tunnel
via excited states.
Thus, the strong decrease of the relaxation rate due to the
(longitudinal) field-detuning of tunnel levels manifests itself in
$\Xnl$ as a distinct quantum contribution, with a sign opposite to the
classical case.
Then, from the signs of the $\Xnl$ vs.\ $\w$ peaks one can estimate if
quantum effects play a role in the dynamics of the studied nanomagnet.
Finally, from the analysis of the angular dependence of $\Xnl(\w)$ we
have estimated a bound for the decoherence time required to attain a
diagonal density matrix due to the phonon-bath coupling.
The so obtained $\tphi$ is much shorter than the level lifetime
$\tau_{0}$ and is the responsible for a fast loss of coherent dynamics
like tunnel oscillations or precession.

%%%%%%%%%%%%%%%%%%%%%%%%%%%%%%%%%%%%%%%%%%%%%%%%%%%%%%%%%%%%
%%%%%%%%%%%%%%%%%%%%   ACKNOWLEDGEMENTS   %%%%%%%%%%%%%%%%%%
%%%%%%%%%%%%%%%%%%%%   ACKNOWLEDGEMENTS   %%%%%%%%%%%%%%%%%%
%%%%%%%%%%%%%%%%%%%%   ACKNOWLEDGEMENTS   %%%%%%%%%%%%%%%%%%
%%%%%%%%%%%%%%%%%%%%   ACKNOWLEDGEMENTS   %%%%%%%%%%%%%%%%%%
%%%%%%%%%%%%%%%%%%%%   ACKNOWLEDGEMENTS   %%%%%%%%%%%%%%%%%%
%%%%%%%%%%%%%%%%%%%%%%%%%%%%%%%%%%%%%%%%%%%%%%%%%%%%%%%%%%%%

\begin{acknowledgments}
This work has been funded by projects MAT$02-0166$ and BFM$2002-00113$
from Ministerio de Ciencia y Tecnolog\'{\i}a and {\sc pronanomag}
from Diputaci\'on General de Arag\'on (Spain).
RLR acknowledges a grant from Consejo Superior de Investigaciones
Cient\'{\i}ficas.
%
%We thank V. Gonz\'{a}lez for her participation in the first
%equilibrium measurements and A. Mill\'an for the sample preparation.
\end{acknowledgments}

%%%%%%%%%%%%%%%%%%%%%%%%%%%%%%%%%%%%%%%%%%%%%%%%%%%%%%%%%%%%
%%%%%%%%%%%%%%%%%%%%%%   APENDICES   %%%%%%%%%%%%%%%%%%%%%%%
%%%%%%%%%%%%%%%%%%%%%%   APENDICES   %%%%%%%%%%%%%%%%%%%%%%%
%%%%%%%%%%%%%%%%%%%%%%   APENDICES   %%%%%%%%%%%%%%%%%%%%%%%
%%%%%%%%%%%%%%%%%%%%%%   APENDICES   %%%%%%%%%%%%%%%%%%%%%%%
%%%%%%%%%%%%%%%%%%%%%%   APENDICES   %%%%%%%%%%%%%%%%%%%%%%%
%%%%%%%%%%%%%%%%%%%%%%   APENDICES   %%%%%%%%%%%%%%%%%%%%%%%
%%%%%%%%%%%%%%%%%%%%%%   APENDICES   %%%%%%%%%%%%%%%%%%%%%%%
%%%%%%%%%%%%%%%%%%%%%%%%%%%%%%%%%%%%%%%%%%%%%%%%%%%%%%%%%%%%

\appendix

%%%%%%%%%%%%%%%%%%%%%%%%%%%%%%%%%%%%%%%%%%%%%%%%%%%%%%%%%%%%
%%%%%%%%%%%%%%%%   ANALYTICAL  EXPRESSIONS   %%%%%%%%%%%%%%%
%%%%%%%%%%%%%%%%   ANALYTICAL  EXPRESSIONS   %%%%%%%%%%%%%%%
%%%%%%%%%%%%%%%%   ANALYTICAL  EXPRESSIONS   %%%%%%%%%%%%%%%
%%%%%%%%%%%%%%%%   ANALYTICAL  EXPRESSIONS   %%%%%%%%%%%%%%%
%%%%%%%%%%%%%%%%   ANALYTICAL  EXPRESSIONS   %%%%%%%%%%%%%%%
%%%%%%%%%%%%%%%%%%%%%%%%%%%%%%%%%%%%%%%%%%%%%%%%%%%%%%%%%%%%

\section{
Analytical expressions for various quantities
}
\label{app:analytic}

In this appendix we derive a number of analytical formulas used in the
discussions of the main text.
First, the equilibrium nonlinear susceptibility of a spin $S$ in the
isotropic limit (from the Brillouin magnetization).
Then, corrections due to finite magnetic anisotropy to the equilibrium
linear and nonlinear susceptibilities in the opposite Ising limit.
Finally, we derive the frequency dependent nonlinear susceptibility
(\ref{X3:debye}); we also analyze the zeroes, extrema, and signs of
the formula for $\Xnl(\w)$.
For simplicity, we omit throughout the appendix unessential constants
like $N_{\rm A}$, $g\mu_{\rm B}$, $k_{\rm B}$, etc.

%%%%%%%%%%%%%%%%%%%%%%%%%%%%%%%%%%%%%%%%%%%%%%%%%%%%%%%%%%%%
%%%%%%%%%%%%%%   ISOTROPIC SUSCEPTIBILITIES   %%%%%%%%%%%%%%
%%%%%%%%%%%%%%   ISOTROPIC SUSCEPTIBILITIES   %%%%%%%%%%%%%%
%%%%%%%%%%%%%%   ISOTROPIC SUSCEPTIBILITIES   %%%%%%%%%%%%%%
%%%%%%%%%%%%%%%%%%%%%%%%%%%%%%%%%%%%%%%%%%%%%%%%%%%%%%%%%%%%

\subsection{
Equilibrium isotropic susceptibilities from the Brillouin magnetization
}
\label{app:Xiso}

For isotropic spins, we can get the equilibrium linear and nonlinear
susceptibilities by expanding the Brillouin function around zero
field.
The isotropic magnetization can be written as ($y=H/T$)
%_____________________________
%
% BRILLOUIN MAGNETISATION: A & B
%_____________________________
\begin{equation*}
%\label{Mz:ab}
%%%%%%%%%%%%%%%%%%
\Mz
=
a\,\cth(a\,y)
-
b\,
\cth(b\,y)
%%%%%%%%%%%%%%%%%%
\;,
\qquad
%%%%%%%%%%%%%%%%%%
a=S+\half
\;,
\quad
b=\half
%%%%%%%%%%%%%%%%%%
\;.
\end{equation*}
%_____________________________
%_____________________________
Then, from the first terms of the small $x$ expansion of the
hyperbolic cotangent, namely $\cth(x)\simeq1/x+x/3-x^{3}/45$, we get
%_____________________________
%_____________________________
\begin{equation*}
\Mz
\simeq
\tfrac{1}{3}\,
(a^{2}\!-\!b^{2})\,
y
-
\tfrac{1}{45}\,
(a^{4}\!-\!b^{4})\,
y^{3}
\stackrel{\mathrm{def}}{\equiv}
\chi_{T}
H
%+\chi^{(2)}H^{2}
+
\chi_{3T}
H^{3}
%%%%%%%%%%%%%%%%%%
\;.
\end{equation*}
%_____________________________
%_____________________________
Using now
$a^{2}-b^{2}=S(S+1)$ and $a^{2}+b^{2}=S(S+1)+\half$,
we finally obtain
%_____________________________
%
% CURIE SUSCEPTIBILITY
%_____________________________
\begin{equation}
\label{chi:curie}
%%%%%%%%%%%%%%%%%%
\chi_{T}
=
\frac{S(S+1)}{3T}
%%%%%%%%%%%%%%%%%%
\;,
\quad
%%%%%%%%%%%%%%%%%%
\chi_{3T}
=
-
\frac{S(S+1)[S(S+1)+\half]}{45\,T^{3}}
%%%%%%%%%%%%%%%%%%
\;.
\end{equation}
%_____________________________
%_____________________________
The first is the celebrated {\em Curie Law} and the latter its sought
generalization for the nonlinear response.
Note that there is not second order $\chi_{2}$ since $\Mz$ is odd in $H$.

%%%%%%%%%%%%%%%%%%%%%%%%%%%%%%%%%%%%%%%%%%%%%%%%%%%%%%%%%%%%
%%%%%%%%%%%%%   ANISOTROPIC SUSCEPTIBILITIES   %%%%%%%%%%%%%
%%%%%%%%%%%%%   ANISOTROPIC SUSCEPTIBILITIES   %%%%%%%%%%%%%
%%%%%%%%%%%%%   ANISOTROPIC SUSCEPTIBILITIES   %%%%%%%%%%%%%
%%%%%%%%%%%%%%%%%%%%%%%%%%%%%%%%%%%%%%%%%%%%%%%%%%%%%%%%%%%%

\subsection{
Finite-anisotropy corrections to the equilibrium Ising susceptibilities
}
\label{app:Xani}

Now we go into the corrections due to finite anisotropy to the
susceptibilities in the opposite Ising limit.
In this regime, for uniaxial anisotropy, only the states $m=\pm S$ are
populated at zero field, and each cluster becomes effectively a
two-level system.
We will consider the effects of finite population of the first
excited levels $m=\pm (S-1)$.
For simplicity, and to compare with known results in the classical
case \cite{gar2000}, we assume the simplest form for the magnetic
anisotropy ${\cal H}=-DS_{z}^{2}$ [cf.\ Eq.~(\ref{Hamanis})].

We start from the statistical-mechanical expressions for the
susceptibilities at zero field \cite{garjonsve2000}:
%_____________________________
%
% EQUILIBRIUM SUSCEPTIBILITIES
%_____________________________
\begin{equation}
\label{chis:eq}
%%%%%%%%%%%%%%%%%%
\chi_{T}
=
\frac{\llangle S_{z}^{2}\rrangle}{T}
%%%%%%%%%%%%%%%%%%
\;,
\qquad
%%%%%%%%%%%%%%%%%%
\chi_{3T}
=
\frac{\llangle S_{z}^{4}\rrangle-3\llangle S_{z}^{2}\rrangle^{2}}{6T^{3}}
%%%%%%%%%%%%%%%%%%
\;.
\end{equation}
%_____________________________
%_____________________________
To get the equilibrium averages $\llangle S_{z}^{k}\rrangle$, we need
first the partition function.
Including the contributions of the ground states, $m=\pm S$, and first
excited states, $m=\pm (S-1)$, we simply have
%_____________________________
%
% PARTITION FUNCTION
%_____________________________
\begin{equation}
\label{Z}
%%%%%%%%%%%%%%%%%%
\Z
=
\sum_{m=-S}^{S}
\e^{-\el_{m}/T}
\simeq
2
\Big(
\e^{-\el_{S}/T}
+
\e^{-\el_{S-1}/T}
\Big)
%%%%%%%%%%%%%%%%%%
\;,
\end{equation}
%_____________________________
%_____________________________
where we have taken into account the zero-field degeneracy
$\el_{-m}=\el_{m}$ (yielding the factor $2$).
For $\el_{m}=-D m^{2}$ the separation between adjacent levels is
$\el_{m}-\el_{m-1}=-D(2m-1)$.
Thus, for the ground-state splitting one has
$\el_{S}-\el_{S-1}=-D(2S-1)=-\Omega_{0}$, which corresponds to
Eq.~(\ref{W0}).
Then, extracting a factor $\e^{-\el_{S}/T}$ in $\Z$, we find the
approximate partition function for our problem
%_____________________________
%
% PARTITION FUNCTION: APPROX.
%_____________________________
\begin{equation}
\label{Zcorr}
%%%%%%%%%%%%%%%%%%
\Z
\simeq
2\,
\e^{-\sigma}
\big(
1
+
\e^{-\Omega_{0}/T}
\big)
%%%%%%%%%%%%%%%%%%
\;,
\end{equation}
%_____________________________
%_____________________________
where we have also introduced the reduced anisotropy barrier
$\sigma=DS^{2}/T$.

Next, we compute the moments
$\langle S_{z}^{k}\rangle\propto\sum_{m}m^{k}\e^{-\el_{m}/T}$
required in Eqs.~(\ref{chis:eq}) within the same approximation:
%_____________________________
%_____________________________
\begin{equation}
%\label{}
%%%%%%%%%%%%%%%%%%
\llangle S_{z}^{k}\rrangle
=
\frac{2}{\Z}\,
S^{k}\,
\e^{-\sigma}
\;
\Big[
1
+
\Big(\frac{S-1}{S}\Big)^{k}
\e^{-\Omega_{0}/T}
\Big]
%%%%%%%%%%%%%%%%%%
\;.
\end{equation}
%_____________________________
%_____________________________
We have considered even $k$, yielding the factor $2$; otherwise the
moments vanish (as they are computed at $H=0$).
Then, dividing by $\Z$ as given by Eq.~(\ref{Zcorr}), we have
%_____________________________
%
% MOMENTS: APPROX.
%_____________________________
\begin{equation}
\label{moments}
%%%%%%%%%%%%%%%%%%
\llangle S_{z}^{k}\rrangle
=
S^{k}\,
\frac
{1+\big(\frac{S-1}{S}\big)^{k}\e^{-\Omega_{0}/T}}
{1+\e^{-\Omega_{0}/T}}
%%%%%%%%%%%%%%%%%%
\;.
\end{equation}
%_____________________________
%_____________________________
Inserting these moments for $k=2,4$ in Eqs.~(\ref{chis:eq}), we get
%_____________________________
%
% EQUILIBRIUM SUSCEPTIBILITIES: APPROX
%_____________________________
\begin{eqnarray*}
%\label{chis:eq}
%%%%%%%%%%%%%%%%%%
\chi_{T}
&=&
\frac{S^{2}}{T}
\frac
{1+\big(\frac{S-1}{S}\big)^{2}\e^{-\Omega_{0}/T}}
{1+\e^{-\Omega_{0}/T}}
\\
%%%%%%%%%%%%%%%%%%
\chi_{3T}
&=&
\frac{S^{4}}{6T^{3}}
\bigg\{
\frac
{1+\big(\frac{S-1}{S}\big)^{4}\e^{-\Omega_{0}/T}}
{1+\e^{-\Omega_{0}/T}}
-
3
\Big[
\frac
{1+\big(\frac{S-1}{S}\big)^{2}\e^{-\Omega_{0}/T}}
{1+\e^{-\Omega_{0}/T}}
\Big]^{2}
\bigg\}
%%%%%%%%%%%%%%%%%%
%\;,
\end{eqnarray*}
%_____________________________
%_____________________________
which are the desired susceptibilities incorporating the effects of
finite population of the first excited levels.

The above results show that the corrections to the Ising limits
$\chi_{T}=S^{2}/T$ and $\chi_{3T}=-S^{4}/3T^{3}$ are functionally
exponential.
This can be seen more explicitly as follows.
Note first that $\beta\equiv\e^{-\Omega_{0}/T}$ is a small parameter in
the considered approximation (for $\Omega_{0}\sim15$\,K and
$T\sim5$\,K, we have $\beta\sim5\cdot10^{-2}$).
Then, introducing $f_{k}\equiv(1-1/S)^{k}$, the moments can be
approximated binomially, $(1+x)^{\alpha}=1+\alpha\, x\cdots$, giving
$\langle S_{z}^{k}\rangle
=
(1+\beta\,f_{k})/(1+\beta)\simeq1-\beta\,(1-f_{k})$.
To illustrate $1-f_{2}=(2S-1)/S^{2}$ gives explicitly
%_____________________________
%
% EQUILIBRIUM SUSCEPTIBILITIES: APPROX: 2
%_____________________________
\begin{equation}
%\label{chis:eq}
%%%%%%%%%%%%%%%%%%
\chi_{T}
=
\frac{S^{2}}{T}
\Big[
1
-
\Big(\frac{2S-1}{S^{2}}\Big)
\,
\e^{-\Omega_{0}/T}
\Big]
\;,
%%%%%%%%%%%%%%%%%%
\end{equation}
%_____________________________
%_____________________________
and a result structurally similar for $\chi_{3T}$.
These exponential corrections are to be compared with the power law
corrections in the {\em classical\/} asymptotic results \cite{gar2000}
%_____________________________
%
% EQUILIBRIUM SUSCEPTIBILITIES:
% APPROX: CLASSICAL
%_____________________________
\begin{equation*}
%\label{chis:eq}
%%%%%%%%%%%%%%%%%%
\chi_{T}
\simeq
\frac{S^{2}}{T}
\Big(
1-\frac{1}{\sigma}
\Big)
%%%%%%%%%%%%%%%%%%
\;,
\qquad
%%%%%%%%%%%%%%%%%%
\chi_{3T}
\simeq
-
\frac{S^{4}}{3T^{3}}
\Big(
1-\frac{2}{\sigma}
\Big)
%%%%%%%%%%%%%%%%%%
\;.
\end{equation*}
%_____________________________
%_____________________________
Thus, while in the classical case the corrections enter as
inverse powers of $\sigma=DS^{2}/T$, for quantum spins they are
exponential in $D/T$, leading to a much faster approach into the Ising
regime as the temperature is lowered.

%%%%%%%%%%%%%%%%%%%%%%%%%%%%%%%%%%%%%%%%%%%%%%%%%%%%%%%%%%%%
%%%%%%%%%%%%%%%   NONLINEAR SUSCEPTIBILITY   %%%%%%%%%%%%%%%
%%%%%%%%%%%%%%%   NONLINEAR SUSCEPTIBILITY   %%%%%%%%%%%%%%%
%%%%%%%%%%%%%%%   NONLINEAR SUSCEPTIBILITY   %%%%%%%%%%%%%%%
%%%%%%%%%%%%%%%%%%%%%%%%%%%%%%%%%%%%%%%%%%%%%%%%%%%%%%%%%%%%

\subsection{
Dynamical nonlinear susceptibilities
}
\label{app:X3}

Finally, we proceed to derive Eq.~(\ref{X3:debye}) for the frequency
dependent nonlinear susceptibility $\Xnlw$ used in this work.
The ordinary nonlinear susceptibility, denoted here $\Xnlo$, is
defined in terms of the third harmonic of the response to an
alternating field.
For classical spins with uniaxial anisotropy a formula for $\Xnlo$ was
derived \cite{gargar2004} from a system of low $T$ balance equations
\cite{garkencrocof99} obtained from Brown Fokker--Planck equation
\cite{bro63}, which reads
%_____________________________
%_____________________________
\begin{equation}
\label{Xnlo}
\Xnlo
=
-
\frac{S^{4}}{3T^{3}}\,
\bigg[
\frac{\dBlo^{4}}{1+3\iu\,\w\tau}
-
\frac{3\iu\,\w\tau\;(\glo\dBlo^{4}+\gtr\dBlo^{2}\dBtr^{2})}
{2(1+\iu\,\w\tau)(1+3\iu\,\w\tau)}
\bigg]
\;.
\end{equation}
%_____________________________
%_____________________________
Here $\dBlo=H_{\|}/H=\cos\psi$ and $\dBtr=H_{\perp}/H=\sin\psi$ are
the direction cosines of the probing field, while $\glo$ and $\gtr$ are
the expansion coefficients of the relaxation rate as
given by Eq.~(\ref{expansionG}),
$\rate
\simeq
\rateo[1+\tfrac{1}{2}(\glo\dBlo^{2}+\gtr\dBtr^{2})\,\xi^{2}]$,
with $\rateo=\rate|_{H=0}$ and $\xi=SH/T$
(these $g$'s differ from those of Ref.~\cite{gargar2004} by a factor
$1/2$).
Accepting the validity of the balance equations for the number of
spins pointing up or down, the derivation of Ref.~\cite{gargar2004}
leading to Eq.~(\ref{Xnlo}) is also valid for quantum spins (with
uniaxial anisotropy).

An alternative nonlinear susceptibility
\cite{wuetal93,schetal95,jonsvehan98}, the one we use in the main
text, can be obtained from the first harmonic of the response in the
presence of a weak static field $H$ as $\chi(\w,H)=\chi_{1}(\w)+3\Xnlw
H^{2}+\dots$ [cf.\ Eq.~(\ref{expansion})].
Then, $\Xnlw$ can be obtained from the first harmonic as
%_____________________________
%
%
%_____________________________
\begin{equation}
%%%%%%%%%%%%%%%%%%
\partial_{H}^{2}\chi\big|_{0}
=
3!\Xnlw
%%%%%%%%%%%%%%%%%%
\;.
\end{equation}
%_____________________________
%_____________________________
Experimentally, as discussed in Sec.~\ref{sec:samples-measurements},
one can measure $\chi$ for several $H$ and get $\Xnlw$ by polynomial
fitting.
Both $\Xnlo$ and $\Xnlw$ coincide in the limit $\w\to0$ with the
thermal equilibrium $\chi_{3T}$.
In addition, as we will see here, they have qualitatively similar
dependences on the damping, $\w$, $T$, angle, etc.
However, $\Xnlw$ presents the experimental advantage of not requiring
high-harmonic detection and processing.

To derive a formula for $\Xnlw$ one can proceed as in
Ref.~\cite{gargar2004} using a system of low $T$ balance equations.
Here, however, we shall present a more direct derivation starting from
the Debye form (\ref{debye}) for the first harmonic response.
Let us write this as $\chi(\w,H)=\chi_{S}+\Dchi/1+\iu\,\w\tau$, where
$\Delta\chi=\chi_{T}-\chi_{S}$.
This one-mode relaxation form is valid in weak enough fields
\cite{ferluibar98,pohsch2000,zuegar2005}, just accounting for the $H$
dependences of the $\chi$'s and $\tau$.
We can take advantage of this to get $\Xnlw$ by simple
differentiation $\Xnlw=\partial_{H}^{2}\chi|_{0}/3!$.
In addition, in the low temperature regime we approximate the
susceptibilities by their Ising limits (this is supported by the
experiments of Sec.~\ref{sec:equilibrium}).
Consistently, the magnetization is given by
$\Mz/S=\dBlo\mathrm{th}(\xi_{\|})$, with $\xi_{\|}=\dBlo\xi$.
Then, we write the Debye law as
%_____________________________
%
% DEBYE LAW
%_____________________________
\begin{equation}
\label{debye:2}
%%%%%%%%%%%%%%%%%%
\chi(\w,H)
\simeq
\frac{\Dchi}{1+\iu\,\w\tau}
%%%%%%%%%%%%%%%%%%
\;,
\qquad
%%%%%%%%%%%%%%%%%%
\Dchi
\simeq
\partial_{H}\Mz
%%%%%%%%%%%%%%%%%%
\;.
\end{equation}
%_____________________________
%_____________________________
This low $T$ approximation entails to disregard
$\chi_{S}\ll\chi_{T}-\chi_{S}$ and
$\partial_{H}^{2}\chi_{S}\ll\chi_{3T}$ \cite{gar2000}, and corresponds
to the $2$-state approximation in the balance equations approach
\cite{gargar2004}.

Based on the above considerations we simply write
%_____________________________
%
% DEBYE LAW
%_____________________________
\begin{equation}
\label{debye:3}
%%%%%%%%%%%%%%%%%%
\chi
=
\partial_{H}\Mz
\cdot
D
%%%%%%%%%%%%%%%%%%
\;,
\qquad
%%%%%%%%%%%%%%%%%%
D
=
\frac{\rate}{\rate+\iu\,\omega}
%%%%%%%%%%%%%%%%%%
\;,
\end{equation}
%_____________________________
%_____________________________
and proceed to differentiate to get
$\Xnlw=\partial_{H}^{2}\chi|_{0}/3!$, using eventually
$\rate=\rateo(1+\QFt\,\xi^{2}/2)$.
To work out the second derivative we use the ``binomial'' formula
$(f\,g)''=f''g+2f'g'+f\,g''$, which results in
%_____________________________
%
% CHI: 2ND DERIVATIVE
%_____________________________
\begin{equation}
%%%%%%%%%%%%%%%%%%
\partial_{H}^{2}
\chi
=
\partial_{H}^{3}\Mz
\cdot
D
+
2\,
\partial_{H}^{2}\Mz
\partial_{H}D
+
\partial_{H}\Mz
\partial_{H}^{2}D
%%%%%%%%%%%%%%%%%%
\;.
\end{equation}
%_____________________________
%_____________________________
Now, using the evenness of the rate on $H$, we have
$\partial_{H}\rate|_{0}=0$, which has two important consequences
%_____________________________
%_____________________________
\begin{equation*}
%%%%%%%%%%%%%%%%%%
\partial_{H}D\big|_{0}
=
0
%%%%%%%%%%%%%%%%%%
\;,
\qquad
%%%%%%%%%%%%%%%%%%
\partial_{H}^{2}D\big|_{0}
=
\frac{\iu\,\w\,\partial_{H}^{2}\rate}{(\rate+\iu\,\w)^{2}}\Big|_{0}
%%%%%%%%%%%%%%%%%%
\;.
\end{equation*}
%_____________________________
%_____________________________
Then, from
$\rate=\rateo(1+\QFt\,\xi^{2}/2)$
and
$\partial_{H}=(S/T)\partial_{\xi}$,
along with
$\partial_{H}\Mz|_{0}=\chi_{T}$
and
$\partial_{H}^{3}\Mz|_{0}=3!\chi_{3T}$,
we finally obtain
%_____________________________
%
% NONLINEAR SUSCEPTIBILITY FROM DEBYE
%_____________________________
\begin{equation}
\label{X3:debye:0}
%%%%%%%%%%%%%%%%%%
\Xnl
=
\frac{\chi_{3T}}{1+\iu\,\w\tau}
+
\frac{\chi_{T}S^{2}}{T^{2}}
\frac{\iu\w\tau \; \QFt}{6(1+\iu\,\w\tau)(1+\iu\,\w\tau)}
%%%%%%%%%%%%%%%%%%
\;,
\end{equation}
%_____________________________
%_____________________________
where now $\tau=1/\rateo$ is the zero-field relaxation time.

To compare with Eq.~(\ref{Xnlo}) for $\Xnlo$, we recall that
$\Mz/S=\dBlo\mathrm{th}(\xi_{\|})$ and use the expansion
$\mathrm{th}\,\xi\simeq\xi-\frac{1}{3}\xi^{3}$, whence
$\partial_{H}\Mz=(S^{2}/T)(\dBlo^{2}-\dBlo^{4}\xi^{2})$
and
$\partial_{H}^{3}\Mz|_{0}=6(-S^{4}\dBlo^{4}/3T^{3})$.
Then, introducing the explicit expression for
$\QFt=\glo\dBlo^{2}+\gtr\dBtr^{2}$, we arrive at [cf.\
Eq.~(\ref{X3:debye})]
%_____________________________
%_____________________________
\begin{equation}
\label{Xnlw}
\Xnlw
=
-
\frac{S^{4}}{3T^{3}}\,
\bigg[
\frac{\dBlo^{4}}{1+\iu\,\w\tau}
-
\frac{\iu\,\w\tau\;(\glo\dBlo^{4}+\gtr\dBlo^{2}\dBtr^{2})}
{2(1+\iu\,\w\tau)(1+\iu\,\w\tau)}
\bigg]
\;.
\end{equation}
%_____________________________
%_____________________________
This result is generic and valid for classical and quantum spins,
inasmuch as the starting Debye $\chi$ provides an adequate
description.
In general, $\glo$ and $\gtr$ will not be given by the classical
result (\ref{Gamma:coeffs:app}), but they incorporate quantum
contributions to the relaxation rate.
%
%% For completeness, we also write the form obtained gathering the
%% direction cosines
%% %_____________________________
%% %_____________________________
%% \begin{equation}
%% \label{Xnlw:ang}
%% \Xnlw
%% =
%% -
%% \frac{S^{4}}{3T^{3}}\,
%% \bigg[
%% \dBlo^{4}
%% \;
%% \frac{1+(1-\glo)\iu\,\w\tau}{(1+\iu\,\w\tau)^{2}}
%% -
%% \frac{1}{2}\gtr\dBlo^{2}\dBtr^{2}
%% \;
%% \frac{\iu\,\w\tau}{(1+\iu\,\w\tau)^{2}}
%% \bigg]
%% \;.
%% \end{equation}
%% %_____________________________
%% %_____________________________
%% In this form the angular dependence is better recognized and, in turn,
%% the contributions of the rate expansion coefficients $\glo$ and
%% $\gtr$.

The expression derived for $\Xnlw$ shows a close structural analogy
with that for the third harmonic nonlinear susceptibility [cf.\
Eq.~(\ref{Xnlw}) with (\ref{Xnlo})].
Indeed, we have kept the factor $(1+\iu\,\w\tau)^{2}$ without squaring
in Eq.~(\ref{Xnlw}) to enhance the analogy; replacing
$\w\tau\to3\w\tau$ both quantities exhibit almost the same frequency
dependece.
Comparison of Eq.~(\ref{Xnlw}) with (\ref{Xnlo}) also reveals similar
dependences on $T$, $\glo$ and $\gtr$, as well as on the angle $\psi$.
This supports our repeated claim about the analogous qualitative
dependences of $\Xnlo$ and $\Xnlw$ and, in turn, our choice of the
first harmonic response on the basis of its experimental convenience.

We conclude this appendix finding extrema and zeroes of $\Xnl(\w)$.
This will help to explote having an analytical expression for the
nonlinear susceptibility when analyzing experiments.
%
%We return to the form without angular coefficients gathered
%
First we normalize Eq.~(\ref{Xnlw}) by the equilibrium value
$\Xnlwt=\Xnlw/\chi_{3T}$:
%_____________________________
%
% CHI3
%_____________________________
\begin{equation}
\label{Xnlw:norm}
%%%%%%%%%%%%%%%%%%
\Xnlwt
=
\frac{1}{1+\iu\,x}
-
\frac{\QF}{2}
\frac{\iu\,x}{(1+\iu\,x)^{2}}
%%%%%%%%%%%%%%%%%%
\;,
\qquad
%%%%%%%%%%%%%%%%%%
x
=
\w\tau
%%%%%%%%%%%%%%%%%%
\;,
\end{equation}
%_____________________________
%_____________________________
where
$\QF=\QFt/\dBlo^{2}=\glo+\gtr(\dBtr/\dBlo)^{2}$.
Multiplying now by the conjugate denominators we readily separate the real
and imaginary parts
%_____________________________
%
% CHI3
%_____________________________
\begin{equation*}
%%%%%%%%%%%%%%%%%%
\Xnlwt'
=
\frac{1}{1+x^{2}}
-
\frac{\QF\,x^{2}}{(1+x^{2})^{2}}
%%%%%%%%%%%%%%%%%%
\;,
\quad
%%%%%%%%%%%%%%%%%%
-\Xnlwt''
=
\frac{x}{1+x^{2}}
+
\frac{\QF}{2}\,
\frac{x(1-x^{2})}{(1+x^{2})^{2}}
%%%%%%%%%%%%%%%%%%
\;.
\end{equation*}
%_____________________________
%_____________________________

Let us first compute where the imaginary part crosses the $\w\tau$
axis (see Fig.~\ref{X3normvsfreq}, bottom panel).
To find this requires to solve $x(1+x^{2})+(\QF/2)x(1-x^{2})=0$.
Appart from $x=0$, one finds the following zero:
%_____________________________
%_____________________________
\begin{equation}
\label{zero:Im}
%%%%%%%%%%%%%%%%%%
x_{\rm z}^{2}
=
\frac{\QF+2}{\QF-2}
%%%%%%%%%%%%%%%%%%
\qquad
\stackrel{|\QF|\gg1}{\leadsto}
\qquad
%%%%%%%%%%%%%%%%%%
x_{\rm z}
\simeq
1
+
\frac{2}{\QF}
%%%%%%%%%%%%%%%%%%
\;.
\end{equation}
%_____________________________
%_____________________________
To get the large $\QF$ approximation we have used the binomial formula
$(1+x)^{\alpha}=1+\alpha\,x\cdots$ twice (to work the denominator and
then to take the square root).
Well, note that for $|\QF|<2$ one has $x_{\rm z}^{2}<0$ and hence the
imaginary part does not cross the $x$ axis.
However, when one {\em finds\/} the crossing, the following criterion
holds: if $x_{\rm z}>1$, then $\QF>0$, whereas for $x_{\rm z}<1$, one
has $\QF<0$.
Thus, inspection of the crossing of the $\w\tau$ axis (below or above
$\w\tau=1$), could provide information on the classical or quantum
character of the spin-reversal dynamics.

Let us now find the extrema $x_{\rm m}$ of the real part $\Xnlwt'$.
Writing $\Xnlwt'=[1+(1-\QF)\,x^{2}]/(1+x^{2})^{2}$ one sees that
$\D\Xnlwt'/\D x=0$ implies $-(1+\QF)+(1-\QF)\,x_{\rm m}^{2}=0$, whence
%_____________________________
%_____________________________
\begin{equation}
\label{maximum:Re}
%%%%%%%%%%%%%%%%%%
x_{\rm m}^{2}
=
\frac{\QF+1}{\QF-1}
%%%%%%%%%%%%%%%%%%
\qquad
\stackrel{|\QF|\gg1}{\leadsto}
\qquad
%%%%%%%%%%%%%%%%%%
x_{\rm m}
\simeq
1
+
\frac{1}{\QF}
%%%%%%%%%%%%%%%%%%
\;.
\end{equation}
%_____________________________
%_____________________________
There is a maximum, or minimum, when $|\QF|>1$: in such case, $x_{\rm
  m}>1$ entails $\QF>0$, whereas $x_{\rm m}<1$ yields $\QF<0$.
We conclude giving the value of $\Xnlwt'$ at its extremum
%_____________________________
%_____________________________
\begin{equation}
\label{maximum:Re:height}
%%%%%%%%%%%%%%%%%%
\Xnlwt'(x_{\rm m})
=
-\frac{(\QF-1)^{2}}{4\QF}
%%%%%%%%%%%%%%%%%%
\quad
\stackrel{|\QF|\gg1}{\leadsto}
\quad
%%%%%%%%%%%%%%%%%%
\Xnlwt'(x_{\rm m})
\simeq
-\frac{\QF}{4}
%%%%%%%%%%%%%%%%%%
\;,
\end{equation}
%_____________________________
%_____________________________
which is the result used in Sec.~\ref{dynamical} to get $\glo$ from
the maximum of $\Xnlwt'$ at $\psi=0$ (then $\QF=\glo$).
Note finally that the signs of the $\Xnlwt'$ peak and of $\QF$ are
{\em always\/} opposite, which can help ascertaining the quantum
contribution to the nonlinear dynamics.

%%%%%%%%%%%%%%%%%%%%%%%%%%%%%%%%%%%%%%%%%%%%%%%%%%%%%%%%%%%%
%%%%%%%%%%%%%%%%%%%%%%   BIBLIOGRAPHY   %%%%%%%%%%%%%%%%%%%%
%%%%%%%%%%%%%%%%%%%%%%   BIBLIOGRAPHY   %%%%%%%%%%%%%%%%%%%%
%%%%%%%%%%%%%%%%%%%%%%   BIBLIOGRAPHY   %%%%%%%%%%%%%%%%%%%%
%%%%%%%%%%%%%%%%%%%%%%   BIBLIOGRAPHY   %%%%%%%%%%%%%%%%%%%%
%%%%%%%%%%%%%%%%%%%%%%   BIBLIOGRAPHY   %%%%%%%%%%%%%%%%%%%%
%%%%%%%%%%%%%%%%%%%%%%%%%%%%%%%%%%%%%%%%%%%%%%%%%%%%%%%%%%%%

%\bibliography{/home/jose/escritos/TEX/jlgarcia}

\begin{thebibliography}{10}
\bibitem{endnote41} These are described by the stochastic
Landau-Lifshitz equation or its Fokker--Planck counterpart for the
probability distribution of spin orientations \cite {bro63}.
\bibitem{endnote43}
For random axes the decrease of $\chi _{T}/\chi _{\protect \rm  iso}$
by the small factor $S^{2}/S(S+1)\simeq 0.91$ is probably compensated
by interaction effects.
\bibitem{endnote44}
For the classical calculations, we use the simplest uniaxial
anisotropy ${\cal H}=-DS_{z}^{2}$ but with a barrier $U=70$\,K, equal
to that provided by Hamiltonian~(\ref {Hamanis}).
\bibitem{endnote45}
The relaxation rate $\Gamma$ should be invariant uppon field
inversion, leading to the absence of odd powers in the field expansion
(\ref {expansionG}); the invariance of $\Gamma$ upon field reflection
through the barrier plane yields the vanishing of a mixed term
$\propto \xi _{\|}\xi _{\perp}$ for uniaxial spins.
\end{thebibliography}

%%%%%%%%%%%%%%%%%%%%%%%%%%%%%%%%%%%%%%%%%%%%%%%%%%%%%%%%%%%%
%%%%%%%%%%%%%%%%%%%%%%%%%%%%%%%%%%%%%%%%%%%%%%%%%%%%%%%%%%%%
%%%%%%%%%%%%%%%%%%%%%%%%%%%%%%%%%%%%%%%%%%%%%%%%%%%%%%%%%%%%
%%%%%%%%%%%%%%%%%%%%%%%%%%%%%%%%%%%%%%%%%%%%%%%%%%%%%%%%%%%%
%%%%%%%%%%%%%%%%%%%%%%%%%%%%%%%%%%%%%%%%%%%%%%%%%%%%%%%%%%%%
%%%%%%%%%%%%%%%%%%%%%%%%%%%%%%%%%%%%%%%%%%%%%%%%%%%%%%%%%%%%
%%%%%%%%%%%%%%%%%%%%%%%%%%%%%%%%%%%%%%%%%%%%%%%%%%%%%%%%%%%%

\end{document}